\documentclass[useAMS,usenatbib,usegraphicx]{mn2e}

\usepackage{amsmath}
\usepackage{amsfonts}
\usepackage{rotating}
\usepackage{color}
\usepackage{pdflscape}
\usepackage{txfonts}
\usepackage{natbib, twoopt}
\usepackage[breaklinks=true]{hyperref} 
\usepackage{soul}
\bibpunct{(}{)}{;}{a}{}{,} 
\usepackage[normalem]{ulem}
\newcommand{\aap}{A\&A}

\DeclareGraphicsExtensions{.pdf,.png,.jpg,.eps}

\title[]{On the mass of the Galactic star cluster NGC~4337 \thanks{Based on observations obtained at the Cerro Tololo Inter-American Observatory, Chile, and at ESO Paranal Observatory under program 292.D-5043.}}

\author[]{Anton F. Seleznev$^{1}$\thanks{anton.seleznev@urfu.ru}, Giovanni Carraro$^{2}$, Roberto Capuzzo Dolcetta$^{3}$,  Lorenzo Monaco$^{4}$, and
\newauthor{Gustavo Baume}$^{5,6}$\\
$^{1}$Astronomical Observatory, Ural Federal University, Mira str. 19, Ekaterinburg, 620002, Russia\\
$^{2}$Dipartimento di Fisica e Astronomia, Universit\'a di Padova, Vicolo Osservatorio 3, I-35122, Padova, Italy\\
$^{3}$Dipartimento di Fisica, Sapienza, Universit\'a  di Roma, P.le A. Moro 5, I-00165 Roma, Italy\\
$^{4}$Departamento de Ciencias Fis\'icas, Universidad Andres Bello, Republica
220, Santiago, Chile\\
$^{5}$Facultad de Ciencias Astron\'omicas y Geof\'isicas (UNLP), Universidad de La Plata (CONICET, UNLP), Paseo del Bosque s/n, La Plata, Argentina\\
$^{6}$Instituto de Astrof\'{\i}sica de La Plata (CONICET, UNLP),
   Paseo del Bosque s/n, La Plata, Argentina}

\begin{document}

\date{Accepted 1988 December 15. Received 1988 December 14; in original form 2015 October 11}


\maketitle

\label{firstpage}

\begin{abstract}
Only a small number of galactic open clusters survives for longer than few hundred million years. Longer lifetimes are routinely explained in term of larger initial masses,
particularly quiet orbits, and  off-plane birth-places.
We derive in this work the actual mass of  NGC~4337, one of the few open clusters in the Milky Way inner disk that managed to survive for
about 1.5 Gyr. We derive its mass in  two different ways.
First, we exploit an unpublished photometric data set in the UBVI passbands to estimate -using star counts- the cluster luminosity profile, and luminosity and mass function, and hence its actual mass
both from the luminosity profile and from the mass function.This data-set is also used to infer crucial cluster parameters, as the cluster half-mass radius and distance.
Second,  we make use of a large survey of cluster star radial velocities to derive dynamical estimates for the cluster mass. Under the assumption of virial equilibrium and neglecting the external gravitational field leads to values for the mass significantly larger than those obtained by mean of observed density distribution or with the mass function but still marginally compatible with the inferred values of the invisible mass in form of both low mass stars  or remnants of high mass stars in the cluster.
Finally, we derive the cluster initial mass by computing the mass loss experienced by the cluster during its lifetime, and adopting the various estimates of the actual mass.

\end{abstract}


\section[]{Introduction}
\noindent
NGC~4337 is an intermediate-age, metal rich, open cluster, that received recently some attention, being a rare
example of an old, metal rich, star cluster located in the inner  regions of the Galactic disk.

The first CCD study   by Carraro et al. (2014a), pointed out the potential interest of this object.  Then, a  spectroscopic
follow up of the cluster red giant clump stars with UVES@VLT
by Carraro et al. (2014b) indicated that NGC~4337 is richer in metals than the Sun, and 1.6 Gyr old. In that work a comparison  was performed with a
typical example of intermediate age, metal rich open cluster,  NGC~752. The comparison is particularly intriguing.
Actually, NGC~752 and NGC~4337 share the same age and metal composition,
although they have very different physical structure. NGC~752 (Twarog et al. 2015) is a star cluster on the brink  of dissolution, as one can judge from  its  main sequence (MS), which is heavily depleted in stars a few magnitudes below the turn off (TO) point. The cluster appears also on maps as a diffuse agglomeration of stars, hardly distinguishable from the general Galactic field. NGC~752 owes its discovery and fortune to its particular present-day location, high over the Galactic plane.
On the other side, NGC~4337 is located
close to the Galactic plane,  but appears as a strong star concentration when compared to the surrounding field.
Its MS does not show any evidence of low-mass star depletion to the limit of actual photometry.
One may wonder that the different dynamical evolution of the two clusters is due to several facts. First of all, they may have formed with very different initial masses,
and for this reason after the very same time, NGC~752 is much more dynamically {\it evolved} than NGC~4337.
Second, and assuming that  they were born with the same initial mass, it might have occurred that the orbits of these two clusters were very different, and NGC~752  experienced more strongly the effects of the Galaxy tidal forces. Finally, the two clusters could have the same mass at birth
and underwent a similar degree of interaction with the Galaxy, but they could have originally a significantly different structure.

In an attempt to cast more light into this topic,
in this work we exploit an unpublished photometric data-set in UBVI , and multi-object spectroscopic observations obtained with GIRAFFE@VLT
to derive an estimate of NGC~4337 present-day mass.  We first estimate the luminous mass using the star density profile and the star luminosity function.
Then we estimate the dynamical mass using the stars velocity dispersion both with he assumption of virial equilibrium
and taking into account possible non-stationarity of the cluster and Galactic gravitational field.  Note that virial equilibrium is a commonly accepted condition for star clusters (Davies et al. 2011, Tofflemire et al. 2014, Geller et al. 2015).
In anticipation of the results, we found that the dynamical (virial) mass is  a factor of 5 larger than the luminous mass, and we discuss  different possible explanations.

The paper is organised as follows.  In Sect.~2 we describe how we collect and reduce the photometric data used in this work. Sect.~3 deals with the spectroscopic observations and reduction.  The distance to NGC~4337 is discussed in Sect.~4. Then Sect.~5 is dedicated to  estimate the cluster center and its radius. This,
together with distance are crucial to estimate the luminous and dynamical mass of the cluster. Sect.~6, 7, and 8 illustrates how we derive the luminous mass from the cluster radial density profile and luminosity and mass function. The derivation of the dynamical mass is instead discussed in Sect.~9. Sect.~10, finally,
summarises our results and provides some discussion.

\section{Photometric observations}
\noindent
We took multiple
UBVI images of NGC~4337 in a 20$\times$20 squared arcmin area  on  2006 March 21, at Cerro Tololo Inter-American Observatory,
using the 1.0m ex-Yalo telescope.  operated by the SMARTS
consortium\footnote{\tt http://http://www.astro.yale.edu/smarts}.
This camera is equipped with an STA~$4064\times4064$
CCD\footnote{\texttt{http://www.astronomy.ohio-state.edu/Y4KCam/detector.html}}
with 15-$\mu$m pixels, yielding a scale of 0.289$^{\prime\prime}$/pixel and a
field-of-view (FOV) of $20^{\prime} \times 20^{\prime}$ at the Cassegrain focus of the telescope.
This FOV is large enough to cover the whole cluster and to sample the surrounding Galactic field.  This is
visible in Fig.~1 where we show CCD image for NGC 4337 field.

In Table~1 we present the log of our observations. All observations were carried out in
photometric, good-seeing conditions. Our \emph{UBVI} instrumental photometric system was defined
by the use of a standard broad-band Kitt Peak \emph{UBVI$_{kc}$} set of
 filters.\footnote{\texttt{http://www.astronomy.ohio-state.edu/Y4KCam/filters.html}}
To determine the transformation from our instrumental system to the standard Johnson-Kron-Cousins
system, and to correct for extinction, each night we observed Landolt's area PG~1047 and  SA~98 (Landolt 1992)
multiple times, and with different air-masses.
Field SA~98 in particular includes over 40 well-observed standard stars, with a good magnitude and color coverage:
$9.5\leq V\leq15.8$, $-0.2\leq(B-V)\leq2.2$, $-0.3\leq(U-B)\leq2.1$.

Basic calibration of the CCD frames was done using the Yale/SMARTS Y4K reduction script
based on the IRAF\footnote{IRAF is distributed
by the National Optical Astronomy Observatory, which is operated by the Association
of Universities for Research in Astronomy, Inc., under cooperative agreement with
the National Science Foundation.} package \textsc{ccdred}, and the photometry was performed
using IRAF's \textsc{daophot} and \textsc{photcal} packages. Instrumental magnitudes were extracted using
the point spread function (PSF) method (Stetson 1987) and adopting  a quadratic, spatially variable
master PSF. Finally, the PSF photometry was aperture-corrected using aperture
corrections measured on bright, isolated stars across  the field.

Aperture photometry was then carried out
for all these stars using the PHOTCAL package. We used transformation
equations of the form:\\

\noindent
u = U -0.879$\pm$0.007 + 0.45$\times$(U-B) -0.016$\pm$0.010 $\times$X     (1)  \\
b = B -2.081$\pm$0.010 + 0.25 $\times$(B-V) +0.132$\pm$0.010 $\times$ X                  (2)\\
v = V -2.139$\pm$0.007 + 0.16$\times$ (B-V) -0.021$\pm$0.006 $\times$X   (3)\\
v = V -2.159$\pm$0.007+ 0.16 $\times$ (V-I$_C$)  +0.001$\pm$0.005 $\times$X   (4)\\
 i =  I$_C$ -1.136$\pm$0.005 + 0.08 $\times$ (V-I$_C$)  -0.016$\pm$0.004 $\times$ X                (5)\\

\noindent
where UBVI$_C$ and ubvi are standard and instrumental magnitudes respectively,
X is the airmass of the observation. We adopted typical values for the
extinction coefficients for CTIO (see Baume et al. 2011). To derive V
magnitudes, we used equation 3 when the B magnitude was available; otherwise we used
equation 4.

World Coordinate System (WCS) header information of each frame was obtained
using the ALADIN tool and  Two-Micron All Sky Survey catalog (2MASS) data (Cutri et al. 2003, Skrutskie et al. 2006). The procedure to
perform the astrometric calibration of our data was explained in Baume et al.
(2009). This allowed us to obtain a reliable astrometric calibration ($\sim$0.12").

We used the Starlink Tables Infrastructure Library Tool Set (STILTS)\footnote{\texttt{http://www.star.bris.ac.uk/~mbt/stilts/}}
to manipulate tables and we cross-correlated our
$UBVI_C$ and $JHK$ 2MASS data. We
obtained then a catalogue with astrometric/photometric information of the
detected objects covering approximately a FOV of 20' x 20' of  cluster
region (as in Fig.~1). The full catalog is made available in electronic
form at the Centre de Donnais Stellaire (CDS) website.

\begin{table}
\tabcolsep 0.1truecm
\caption{$UBVI$ photometric observations of NGC~4337 and Landolt standard stars on  Mar 21, 2006.}
\begin{tabular}{cccc}
\hline
\noalign{\smallskip}
Field & Filter & Exposures (s) & airmass (X)\\
\noalign{\smallskip}
\hline
                                    NGC~4337   & \textit{U}  & 10, 30, 200, 1800     & 1.19$-$1.25\\
                                                          & \textit{B}  & 7, 30, 100, 900       & 1.15 $-$1.25\\
                                                         & \textit{V}   & 5, 30, 100, 700        & 1.16$-$1.25\\
                                                         & \textit{I}    & 5, 30, 100, 600        & 1.17$-$1.25\\
                                   SA~101       & \textit{U}  & 2x400                   & 1.20$-$1.24\\
                                                        & \textit{B}  & 2x200                   &  1.19$-$1.22\\
                                                        & \textit{V}  & 2x150                   &  1.19$-$1.22\\
                                                        & \textit{I}    & 2x130                    &  1.20$-$1.23\\
                                   PG~1047     & \textit{U}  & 400                    & 1.20\\
                                                       & \textit{B}  &200                    &  1.18\\
                                                       & \textit{V}  &  150                     &  1.19\\
                                                       & \textit{I}    & 130                     &  1.19\\
                                   SA~104    & \textit{U}  & 200, 400                    & 1.15$-$1.16\\
                                                       & \textit{B}  & 90, 200                    &  1.15$-$1.16\\
                                                       & \textit{V}  & 70,150                    &  1.15$-$1.16\\
                                                       & \textit{I}    & 60, 130                    &  1.15$-$1.16\\
                                  SA~107   & \textit{U}  & 2x200                         & 1.15$-$1.17\\
                                                       & \textit{B}  & 2x90                      &  1.16$-$1.17\\
                                                       & \textit{V}  &  2x70                     &  1.16$-$1.17\\
                                                       & \textit{I}    & 2x60                      &  1.15$-$1.17\\
\hline
\end{tabular}
\end{table}

\begin{table}
\tabcolsep 0.5truecm
\caption{Completeness factors, expressed as percentages, as a function of $V$ magnitude
using $\Delta V = 0.5$ bins, inside the circle of 5.6 arcmin radius of the centre and
outside this circle. The completeness error is $0.5\%$}
\begin{tabular}{ccc}
\hline
Magnitude bin   & Cluster region & Field region \\
\hline
$\le$16.75      &   100.0    &   100.0  \\
16.75-17.25     &    96.7    &    96.9  \\
17.25-17.75     &    97.9    &    98.0  \\
17.75-18.25     &    97.5    &    97.5  \\
18.25-18.75     &    97.0    &    96.8  \\
18.75-19.25     &    96.5    &    96.2  \\
19.25-19.75     &    95.2    &    95.4  \\
19.75-20.25     &    90.9    &    91.9  \\
20.25-20.75     &    78.4    &    80.5  \\
20.75-21.25     &    54.8    &    57.3  \\
21.25-21.75     &    28.4    &    29.2  \\
21.75-22.25     &    10.8    &    10.5  \\
22.25-22.75     &     4.2    &     4.3  \\
22.75-23.25     &     2.9    &     3.2  \\
23.25-23.75     &     2.5    &     2.8  \\
23.75-24.25     &     1.1    &     1.4  \\
\hline
\end{tabular}
\end{table}

\subsection{Completeness}

To estimate the photometric completeness of our data, we carried out
several artificial-star experiments (see Carraro et al. 2005). To this
aim, we generated 50 new images by adding artificial stars
in random positions in the original images, which we then reduced using the same
set of parameters. The mentioned 50 new images correspond to
the amount of experiments for each selected long exposure pass-band (V and I) to evaluate the
completeness factors (CF) and their corresponding errors. To preserve the stellar crowding of the original
images, the amount of added stars for each experiment was 10$\%$ of the total stars, and followed their
colour and luminosity distributions.
The completeness factors (CF) were then computed  as the ratio between the number
of artificial stars recovered by the PSF photometry procedure and the number of
artificial stars added. To minimise the error of the computed ratios, we first
added the amount of all the new stars and the amount of the detected ones for
all the experiments. Out of the 50 experiments per image we estimated also a completeness
error of 0.5$\%$.
The whole procedure was applied to estimate the photometric completeness in two areas of the images.
One within 5.6 arcmin from the cluster centre, to estimate the cluster completeness,
and the other outside this circle, to estimate the field completeness.
Values of CF are listed in Table~2 for 0.5 magnitude bins of V. One can notice that there is
tendency of the cluster completeness to be lower than the field completeness. This is excepted, given
the larger crowding of the cluster area. However, the difference is not large, and in some cases the completeness values are compatible
within the estimated error.
This is probably due to the fact that the cluster is not particularly crowded, and it is projected toward a rich stellar field.

\begin{figure}
\includegraphics[width=\columnwidth]{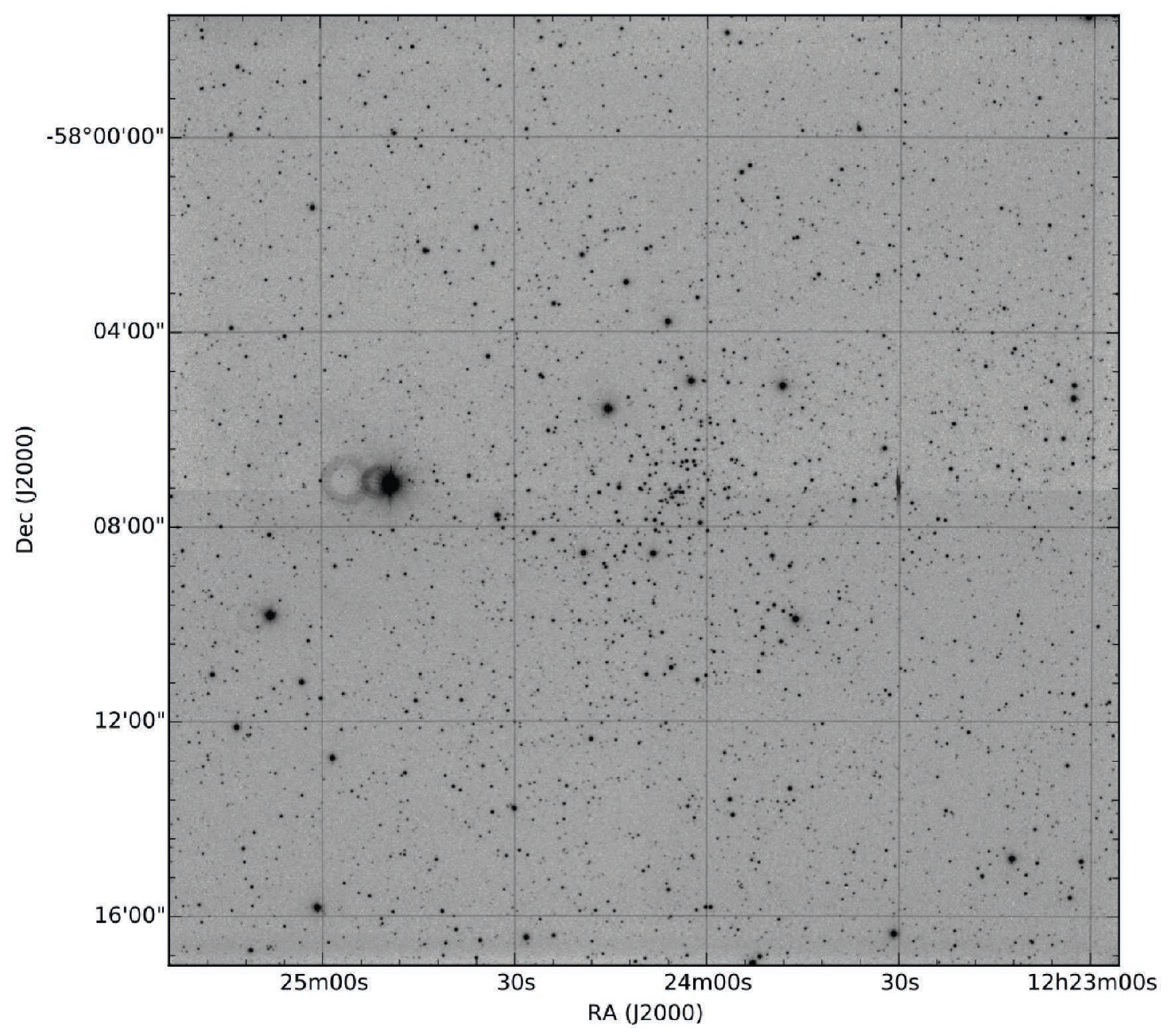}
\caption{A 120 sec frame in B for NGC~4337. North is up, East to the left. The field is 20 arcmin on a side.}
\end{figure}

\section{Spectroscopic observations}
\noindent
We observed red-clump, turn-off (TO) and main sequence (MS) stars belonging to
NGC4337, using the multi-object, fiber-fed spectroscopy facility FLAMES, mounted
at the UT2 telescope  of the VLT. 113 stars were observed with one single plate
configuration on March 30, 2014 for an integration time of 2775s. Observations
were conducted in combined medusa mode, and both the UVES and GIRAFFE
spectrographs were employed. Seven red clump stars were observed with red arm of
UVES, as already reported in Carraro et al. (2014b), while 106 red-clump, TO and
MS stars were observed using GIRAFFE. Sixteen GIRAFFE fibres were allocated to
sky position for sky subtraction. UVES and GIRAFFE spectra have wavelength
coverage and resolution of $\lambda$=4760---6840$\AA$, R=47,000 (UVES, central
wavelength 5800$\AA$) and  $\lambda$=6470---6790$\AA$, R=17,000 (GIRAFFE setup
HR15N).

Data reduction was performed using the ESO CPL based FLAMES-UVES (v.5.3.0) and
FLAMES-GIRAFFE (v2.11.1s)
pipelines\footnote{\url{http://www.eso.org/sci/software/pipelines/}}. The
GIRAFFE fibres allocated to sky holes were finally median combined and
subtracted from the GIRAFFE stellar spectra. Heliocentric correction was
computed using the standard IRAF\footnote{IRAF is distributed by the National
Optical Astronomy Observatory, which is operated by the Association of
Universities for Research in Astronomy (AURA) under a cooperative agreement with
the National Science Foundation.} task {\it rvcorrect}.

We used the IRAF task {\it fxcor} to measure the stellar radial velocity by
cross-correlation with  synthetic spectra of the Coelho et al. (2005)
collection. The synthetic spectra were broadened to the instrumental resolution
before the cross-correlation. We used synthetic spectra having effective
temperatures and surface gravities of $T_{eff}$=4750\,K and 6250\,K and
log\,g=2.5 and 4.0 for red-clump and TO, MS stars, respectively. Individual
measured radial velocities and the corresponding formal {\it fxcor} errors are
reported in Carraro et al. (2014) for the stars observed with UVES and all the
stars have radial velocities consistent with cluster membership. From these
seven stars only, with a jackknife bootstrapping technique (Lupton 1993), we
derive a mean cluster radial velocity and velocity dispersion of
$\langle v_r \rangle$ =-17.76$\pm$0.33\,km\,s$^{-1}$ and $\langle \sigma \rangle$=0.78$\pm$0.61\,km\,s$^{-1}$.

The GIRAFFE sample includes four stars selected in the red-clump CMD region, and
102 stars in the TO, MS region.  Given the expected significant field
contamination in the TO, MS region of the colour-magnitude diagram (CMD), in
order to select likely cluster members, we selected stars with formal error from
the {\it fxcor} fit lower than 5\,km\,s$^{-1}$ and we applied a 2-sigma clipping
rejection to this sample. With this selection. we ended up with 45 likely
cluster members (see Table~3).\\

\noindent
To make our selection more solid and convincing, we show a multi-panel plot in Fig.~2.
In the left-bottom panel  the target stars observed with GIRAFFE (symbols) are plotted on top of the V vs V-I color magnitude diagram of NGC 4337 (dots). Stars used in the calculation of the cluster mean radial velocity and velocity dispersion (see text) are marked as filled triangles, while large open circles are stars excluded by the above selection.
In the left-top panel the formal error on the radial velocity measured with the favor cross-correlation task is plotted against the stellar V magnitude. The expected trend of larger error with fainter magnitudes is evident. The cut at errors larger than 5 km/s applied is shown as a continuous line. The dotted line represents a possible alternative selection on errors, which would retain stars in the lower envelope described by the distribution. Four additional stars would excluded by this selection (see below).
In the right-bottom panel we show the measured radial velocities against the distance from the cluster center, adopted at (RA, Dec)=($186^o.0$, $-58^o.123$). The vertical dashed lines are $2-\sigma$ limits from the cluster mean, after applying a $2-\sigma$-clipping rejection. Open circles in the region delimited by the two dashed lines are stars excluded from the selection due to their RV errors larger than 5 km/s.
The right-middle panel shows the measured radial velocities against the stellar V magnitude.
Finally, the right-top panel shows a histograms of all the radial velocities (dotted line) measured from GIRAFFE spectra and of the radial velocities of the stars retained above as radial velocity members (continuous line) for calculating the cluster mean radial velocity and velocity dispersion. \\

\noindent
With the help of this figure, we can  argue that
the lower envelope of the mean trend reaches about 5 km/s at V=17, which is the faintest magnitudes we reached.
There are, however, several outliers with respect to the mean trend. Most of them are excluded with the cut in error we applied.\\
A, perhaps, better motivated selection would apply a cut in RV error which scale with magnitude. We found that a proper relation would be to accept stars having RV errors lower than:

$err(RV) < 1.14\times V -14.36$\\

\noindent
By applying  this criteria, four additional stars are excluded, before applying the 2-sigma clipping procedure,  at magnitude between V=14.5-15.5. These four stars have, however, errors lower than 5 km/s and are not so evidently discrepant from the bulk of the mean trend.
With this new selection, we would retain 41 stars and obtain:\\

$<RV>$=-17.80$\pm$0.26\,km\,s$^{-1}$ and $\sigma$=1.67$\pm$0.13\,km\,s$^{-1}$\\

which is totally consistent with applying the cut on errors at 5 km/s.
Finally, if not cut on the formal {\it fxcor} error is applied before the 2-sigma clipping procedure, we end up retaining 53 stars and obtaining:\\

$<RV>$=-17.93$\pm$0.21\,km\,s$^{-1}$ and $\sigma$=1.54$\pm$0.10\,km\,s$^{-1}$\\

\noindent
We believe therefore that using the formal  {\it fxcor} is a good choice, and will use the original 45 members in the following of the paper.\\

\noindent
By applying again the jackknife resampling technique, we
finally obtain  a mean cluster radial velocity and velocity dispersion of
$\langle v_r \rangle$=-17.78$\pm$0.28\,km\,s$^{-1}$ and $\langle \sigma \rangle$=1.64$\pm$0.13\,km\,s$^{-1}$.
These values are formally consistent (within 2-sigma) with the results from the
UVES sample (Carraro et al. 2014b) . While the mean cluster radial velocity is practically identical for
the two samples, the radial velocity dispersion obtained from the GIRAFFE sample
is significantly larger and seemingly with a smaller error. The UVES sample is
however  significantly smaller in size (7 against 45 stars). We will consider in the following the velocity dispersion
derived from GIRAFFE as representative of the cluster dispersion. The full table with all the 113 radial velocity
measurements will be made available electronically.

\begin{table*}
\label{tab4}
\tabcolsep 0.1truecm
\caption{Radial velocities. ID are from Carraro et al. (2014a). The last column reports the formal fxcor error.}
\begin{tabular}{ccccccc}
\hline
       ID     & V   & (V-I)    &RA(2000.0) & Dec(2000.0) & RV &   error\\
              & mag & mag  & $hh:mm:ss.sss$ & $dd:mm:ss.ss$& km/s & km/s \\
\hline
$NGC4337\_000169$ & 14.31 & 0.99 & $12:23:51.895$ & $-58:10:58.94$ & -16.7629 & 0.787\\
$NGC4337\_000182$ & 14.43 & 0.90 & $12:23:56.702$ & $-58:05:03.12$ & -15.6799 & 0.824\\
$NGC4337\_000189$ & 14.47 & 0.89 & $12:24:06.994$ & $-58:10:11.03$ & -16.8177 & 1.048\\
$NGC4337\_000197$ & 14.53 & 0.95 & $12:24:13.800$ & $-58:07:53.47$ & -15.5483 & 1.106\\
$NGC4337\_000202$ & 14.54 & 0.90 & $12:24:27.730$ & $-58:07:16.07$ & -18.3345 & 3.651\\
$NGC4337\_000218$ & 14.61 & 0.91 & $12:23:28.104$ & $-58:07:30.72$ & -16.2074 & 0.857\\
$NGC4337\_000256$ & 14.76 & 0.68 & $12:24:03.012$ & $-58:06:50.26$ & -19.6265 & 1.083\\
$NGC4337\_000262$ & 14.81 & 0.94 & $12:24:44.398$ & $-58:05:47.11$ & -15.7375 & 4.443\\
$NGC4337\_000269$ & 14.85 & 0.86 & $12:24:06.979$ & $-58:07:57.04$ & -17.2737 & 3.113\\
$NGC4337\_000283$ & 14.90 & 0.88 & $12:24:09.497$ & $-58:07:51.64$ & -18.5341 & 2.503\\
$NGC4337\_000291$ & 14.92 & 0.85 & $12:24:01.858$ & $-58:06:25.27$ & -20.3456 & 1.354\\
$NGC4337\_000321$ & 15.02 & 0.87 & $12:23:43.987$ & $-58:07:57.40$ & -15.7687 & 1.266\\
$NGC4337\_000329$ & 15.04 & 0.90 & $12:24:05.244$ & $-58:06:42.16$ & -18.8798 & 3.374\\
$NGC4337\_000370$ & 15.15 & 0.87 & $12:24:34.867$ & $-58:09:08.68$ & -14.6227 & 2.360\\
$NGC4337\_000372$ & 15.16 & 0.77 & $12:24:00.091$ & $-58:11:03.80$ & -20.8658 & 1.395\\
$NGC4337\_000433$ & 15.31 & 0.89 & $12:23:41.124$ & $-58:09:58.46$ & -15.7845 & 0.812\\
$NGC4337\_000443$ & 15.33 & 0.88 & $12:24:00.936$ & $-58:07:31.91$ & -18.8097 & 2.286\\
$NGC4337\_000459$ & 15.38 & 0.86 & $12:24:04.010$ & $-58:05:31.88$ & -19.1991 & 1.498\\
$NGC4337\_000460$ & 15.38 & 0.90 & $12:24:08.304$ & $-58:09:34.63$ & -16.6502 & 1.188\\
$NGC4337\_000474$ & 15.41 & 0.91 & $12:23:37.354$ & $-58:07:12.47$ & -18.0435 & 0.898\\
$NGC4337\_000484$ & 15.44 & 0.81 & $12:23:50.407$ & $-58:09:43.06$ & -19.0432 & 1.695\\
$NGC4337\_000539$ & 15.59 & 0.86 & $12:24:21.283$ & $-58:07:39.11$ & -19.9470 & 2.048\\
$NGC4337\_000610$ & 15.73 & 0.89 & $12:24:21.936$ & $-58:11:57.95$ & -20.0636 & 1.819\\
$NGC4337\_000682$ & 15.89 & 0.87 & $12:23:34.356$ & $-58:09:57.53$ & -17.1483 & 0.704\\
$NGC4337\_000694$ & 15.91 & 0.87 & $12:23:52.399$ & $-58:10:15.31$ & -19.5864 & 2.160\\
$NGC4337\_000756$ & 16.03 & 0.92 & $12:24:08.902$ & $-58:07:14.45$ & -16.0065 & 3.533\\
$NGC4337\_000778$ & 16.07 & 0.90 & $12:24:31.361$ & $-58:07:51.35$ & -15.6437 & 2.530\\
$NGC4337\_000839$ & 16.18 & 0.91 & $12:24:42.842$ & $-58:09:14.98$ & -18.8237 & 2.170\\
$NGC4337\_000853$ & 16.21 & 0.95 & $12:24:32.122$ & $-58:03:55.80$ & -16.7957 & 2.156\\
$NGC4337\_000881$ & 16.25 & 0.90 & $12:24:08.914$ & $-58:05:31.78$ & -18.6547 & 3.175\\
$NGC4337\_000918$ & 16.29 & 0.93 & $12:24:01.690$ & $-58:05:35.63$ & -16.7409 & 1.717\\
$NGC4337\_001010$ & 16.41 & 0.94 & $12:23:57.686$ & $-58:04:54.88$ & -19.9451 & 2.400\\
$NGC4337\_001058$ & 16.48 & 0.97 & $12:24:21.312$ & $-58:08:10.18$ & -18.8888 & 3.001\\
$NGC4337\_001090$ & 16.51 & 0.96 & $12:23:46.778$ & $-58:09:39.53$ & -19.7798 & 2.552\\
$NGC4337\_001108$ & 16.54 & 0.95 & $12:24:28.673$ & $-58:05:58.85$ & -20.4966 & 3.097\\
$NGC4337\_001210$ & 16.68 & 0.97 & $12:23:37.944$ & $-58:10:32.02$ & -15.6153 & 1.449\\
$NGC4337\_001264$ & 16.74 & 1.00 & $12:24:19.078$ & $-58:02:55.32$ & -17.5050 & 2.688\\
$NGC4337\_001271$ & 16.75 & 0.95 & $12:24:00.290$ & $-58:05:01.61$ & -17.1769 & 2.417\\
$NGC4337\_001301$ & 16.78 & 1.01 & $12:23:43.613$ & $-58:12:26.68$ & -18.1352 & 1.736\\
$NGC4337\_001403$ & 16.90 & 0.99 & $12:23:23.040$ & $-58:06:35.75$ & -16.7813 & 3.057\\
$NGC4337\_001408$ & 16.91 & 0.99 & $12:23:36.566$ & $-58:08:20.33$ & -18.9967 & 2.245\\
$NGC4337\_001461$ & 16.96 & 1.01 & $12:23:39.266$ & $-58:06:56.66$ & -16.6959 & 4.066\\
$NGC4337\_100110$ & 13.91 & 1.32 & $12:23:57.209$ & $-58:07:14.45$ & -16.1309 & 0.554\\
$NGC4337\_100124$ & 14.02 & 1.32 & $12:24:31.786$ & $-58:08:02.11$ & -17.6769 & 0.544\\
$NGC4337\_100130$ & 14.06 & 1.36 & $12:24:04.898$ & $-58:05:09.85$ & -18.2653 & 0.500\\
\hline
\end{tabular}
\end{table*}

\section{Cluster distance}
\noindent
The distance to NGC~4337 is essential to derive the cluster linear radius, and to convert light into mass.
In Carraro et al. (2014b)  it was established that the cluster is metal rich, with [Fe/H]=+0.12$\pm$0.05,
and with an age of 1.6$\pm$0.1 Gyr.  We refine here the cluster distance and reddening by fitting the CMD
of NGC~4337 using the radial velocity members only (see Table~3). This is shown in Fig.~3. The cluster MS for members only is very clearly defined,
with the typical curvature of intermediate age cluster in the TO region. The TO is located at V = 15 mag.
A clump spread in color is located at the mean magnitude V = 13.9. Two stars bluer than (V-I) $\sim$ 0.8 mag
can be cluster blue stragglers.

We find that an isochrone
of 1.5 Gyr (Bressan et al. 2012) better reproduces the shape of the TO and the magnitude of the clump. This convincing fit yields
a reddening E(V-I) = 0.385$\pm$0.005, and an apparent visual distance modulus (m-M)$_V$ = 12.72$\pm$0.02 mag, where
uncertainties are estimated by visual inspection. In other words we eyeballed the location of the isochrone with respect to the member-stars
by displacing it along the vertical and horizontal direction iteratively, until an acceptable fit could not be found.
This implies that the cluster distance is 2.2$\pm$0.1 kpc
from the Sun, slightly lower than Carraro et al. (2014b) estimate.

\begin{figure}
\includegraphics[width=\columnwidth]{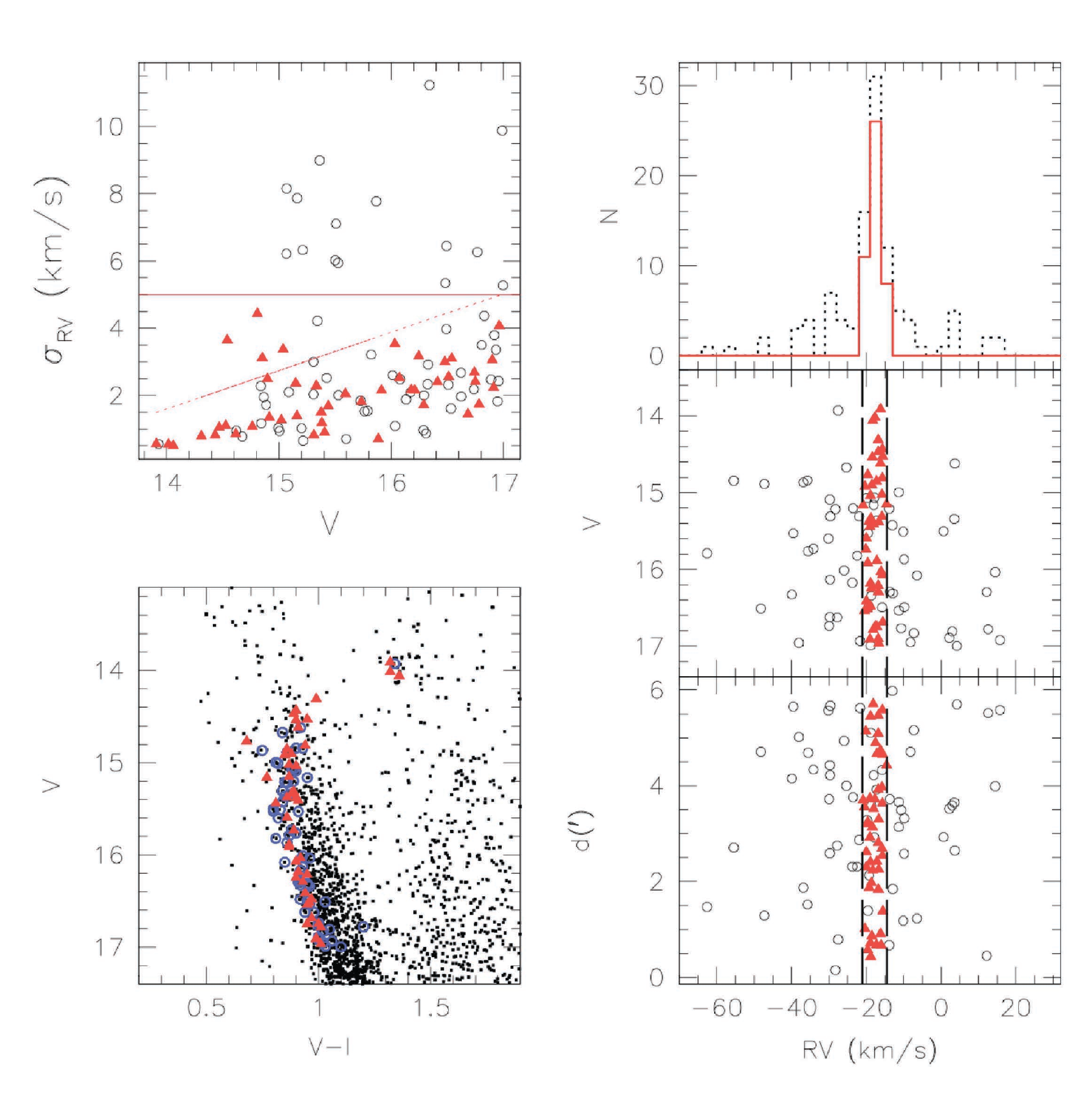}
\caption{Spectroscopic membership assessment. See text for details.}
\end{figure}

\begin{figure}
\includegraphics[width=\columnwidth]{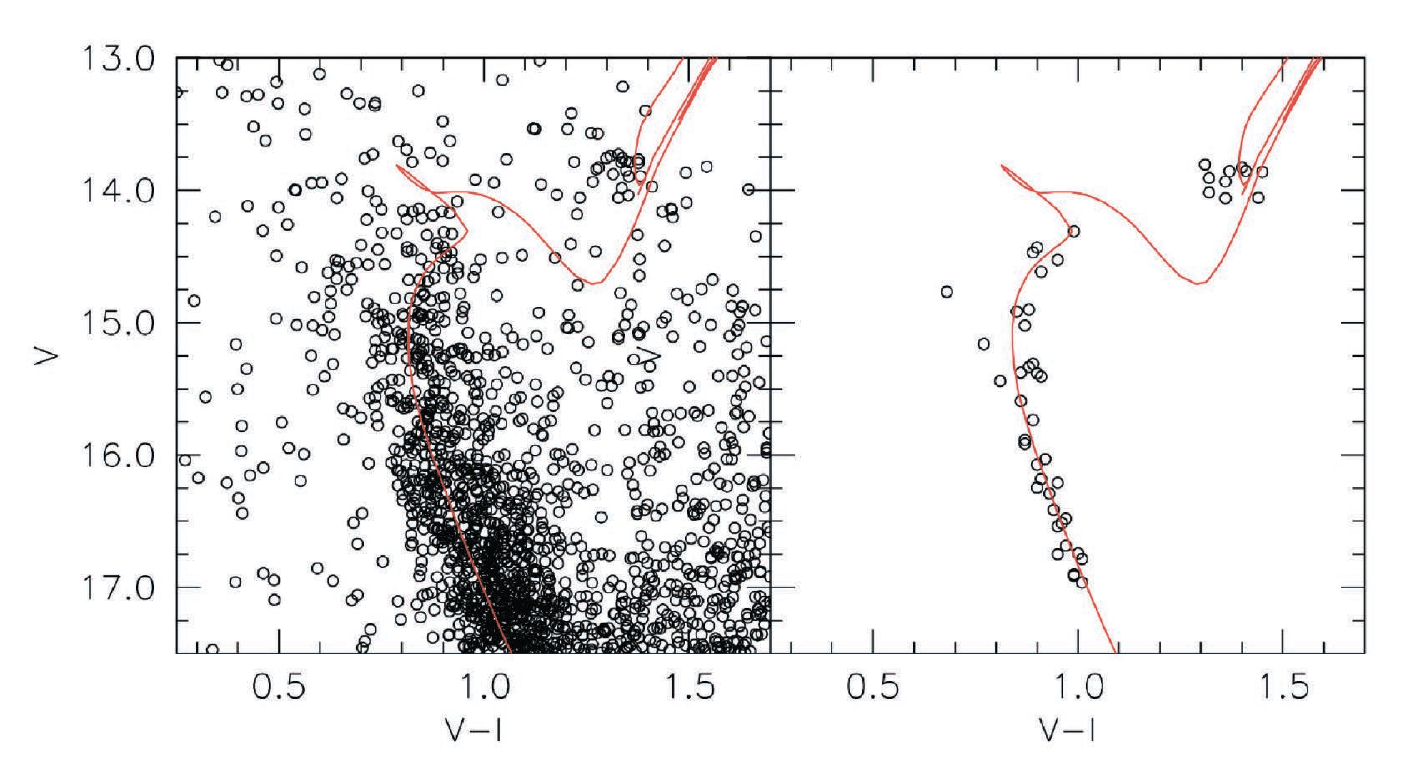}
\caption{Distance determination. The left panel shows all the stars detected in V and I, while the right panel
shows RV members only. The isochrone is for an age of 1.5 Gyr and a metallicity Z=0.025}
\end{figure}

\section{Surface density maps and cluster center}
\noindent
To determine the coordinates of the cluster centre, and to study the cluster 2-dimensional (2D) structure,
we derived surface density maps (2D density distributions)  by using
a kernel estimator (see, for example, Seleznev et al. 2010, and Carraro \& Seleznev  2012,
and the  detailed description of this method in Silverman 1986). Density maps were derived at
varying the limiting magnitude. We warn the reader that star counts cannot be computed in a strip
a kernel halfwidth $h$ wide close to the border of the field,
to prevent under-sampling.

Fig.~4 (left panel) shows the surface density map, centred at RA=$12^h24^m04^s$ and DEC=$-58^\circ07'24''$.
Both $x$ and $y$ coordinates are in arcmin,  with $x$ increasing towards the East, and
$y$ towards the North.
Only stars with $V\leq16$ mag are considered, and we adopted 3 arcmin as kernel
halfwidth. Density units are $arcmin^{-2}$, density values are shown for thick contour lines. This map was chosen
as the best representation of the cluster star distribution,  taking into account the
colour-magnitude diagram, and with  the sole purpose of deriving an estimate of the cluster center.
The cluster centre position was determined in two ways. First, the contour line corresponding
to the density value of 7 $arcmin^{-2}$ was approximated with an ellipse in the polar coordinates following to Pancino et al. (2003).
Then the centre position was determined as the mean value of coordinates of the contour line
in 30-degree-intervals of position angles. Second, the centre coordinates were determined as $x$ and $y$
coordinates of maxima of corresponding linear density functions, obtained
by the kernel estimation with values of the kernel halfwidth of 1 and 3 arcmin. The mean coordinates
of the cluster centre by both methods are $x=-0.34\pm0.15$ arcmin and $y=0.02\pm0.13$ arcmin.
Finally the cluster center was adopted as RA=186$^o$.00594 and DEC=-58$^o$.12300 with uncertainty of about
10 arcseconds.
The robustness of this centre position was confirmed by radial surface density profiles (see below; if the centre location is
determined incorrectly the density profile can have the minimum at the centre).

The middle panel in Fig.~4 shows the surface density map with the same parameters as in left panel,
but for fainter stars ($V \leq 20$ mag). The cluster appears to be stretched from  North-East to  South-West. This distortion  is supported by
surface density map derived with the same method using  2MASS data, and shown in the right panel.
For this case, the adopted kernel halfwidth
is 2 arcmin and the limiting magnitude $J=16$ mag. Elongation of an open cluster can arise, for example, due
to tidal action of the Galaxy. It was studied in numerical experiments (see, for example, Kharchenko et al. (2009) and
Chumak et al. (2010)) and revealed in real clusters (see, for example, Davenport \& Sandquist (2010)). Thus, we have
every reason to consider stars forming elongations of NGC 4337 as possible cluster members. The white
box in the right panel of Fig.~4 indicates the field covered by our optical photometry. Due to the elongated shape of the
cluster, the only possible regions for estimating the field density is in the North-West and the South-East corners of
this field. We remind the reader that these field regions are necessary for deriving the luminosity function of the cluster and its mass, as discussed in the following.

\begin{figure*}
\includegraphics[height=5.8truecm]{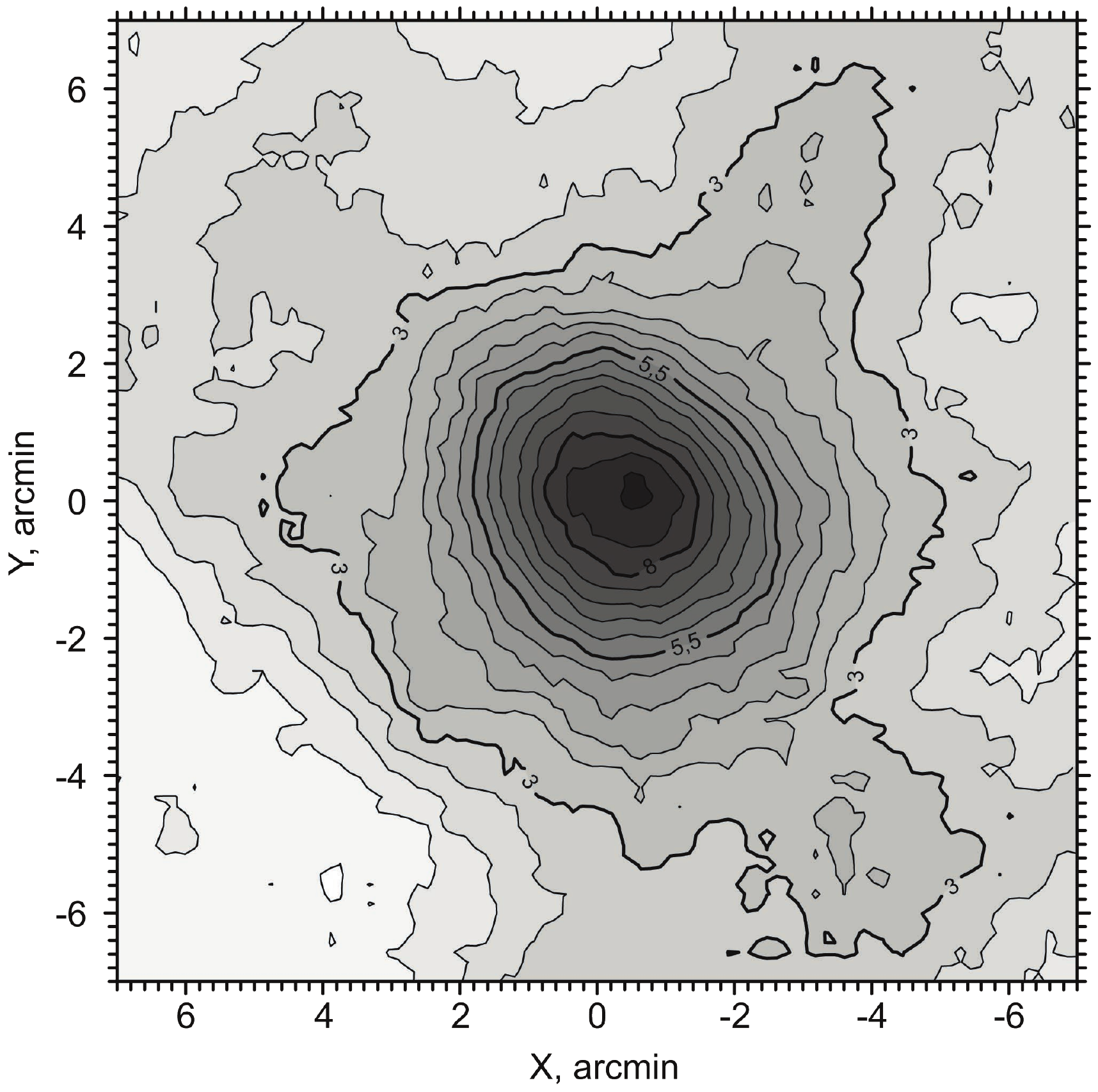}
\includegraphics[height=5.8truecm]{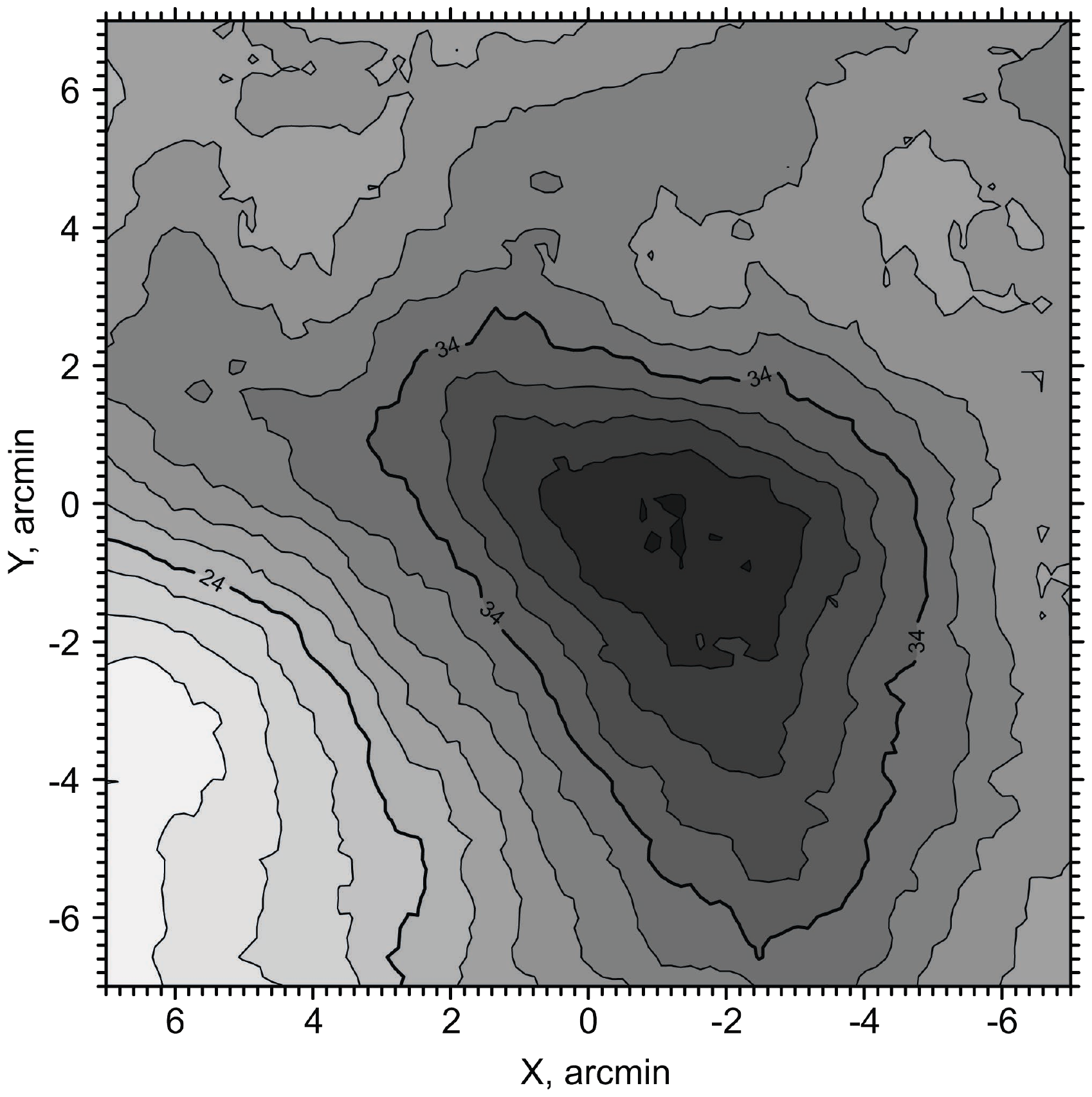}
\includegraphics[height=5.8truecm]{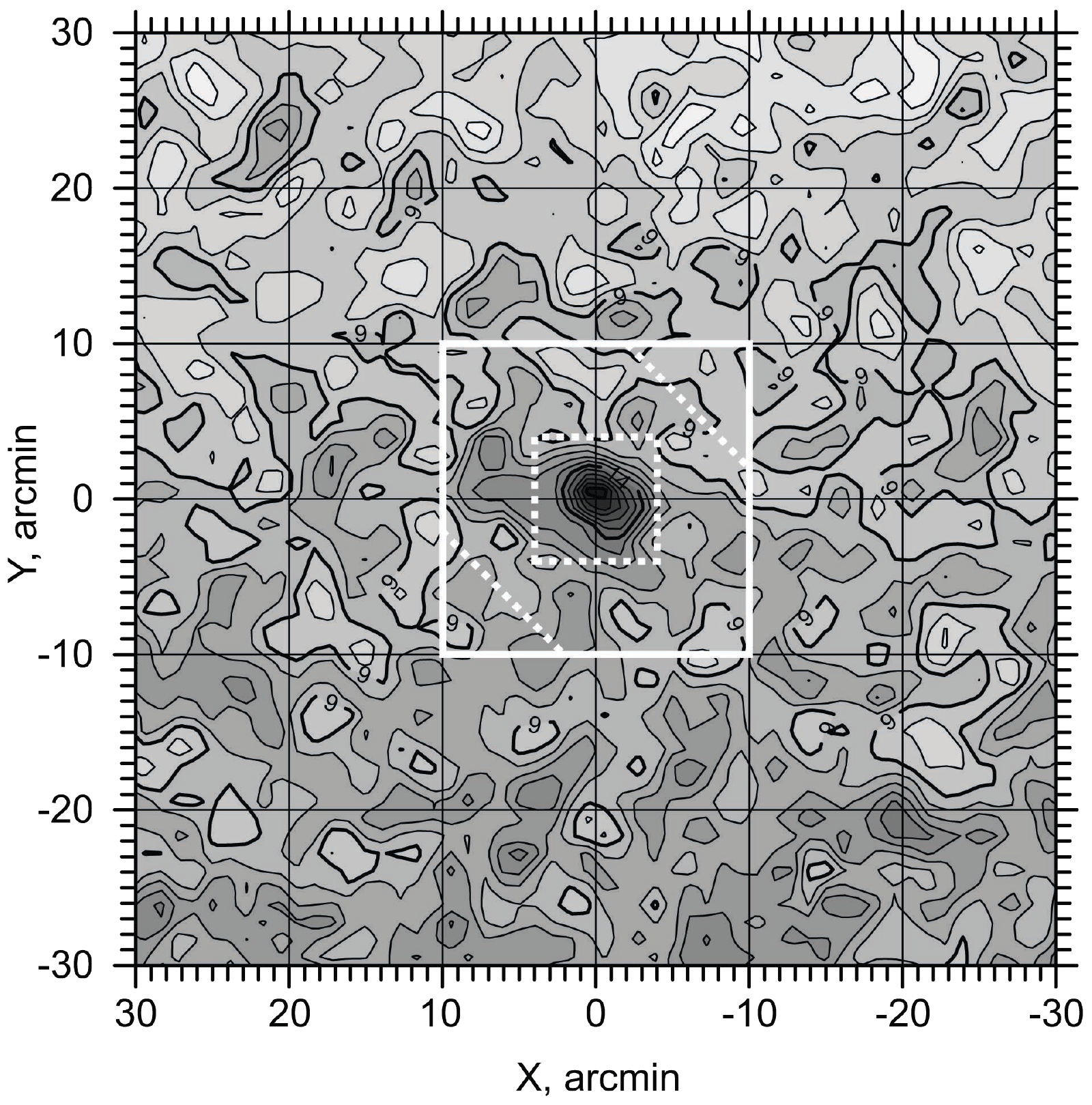}
\caption{{\bf Left panel}: The map of surface density of the cluster for  stars with $V \leq16$ mag. The density units are
$arcmin^{-2}$.
{\bf Middle panel}: the map of surface density of the cluster with stars having $V\leq 20$ mag.
{\bf Right panel}:  the map of surface density of the cluster with  stars having  $J \leq16$ mag,
and derived from 2MASS. The rectangle of white solid line shows the field covered by our optical data. The rectangle of white dotted
line shows the cluster field for LF derivation, while the white dotted lines in the corners of white-solid-lines
rectangle shows the comparison field for LF derivation.}
\end{figure*}

\section{Radial density profiles and cluster mass}
\label{raddenprof}
The cluster radial surface density profiles $F(r)$ were derived using
the cluster centre coordinates obtained  in the previous Sect., and for different limiting
magnitudes $V_{lim}$. The kernel estimator method was used (Merritt \& Tremblay 1994, Seleznev 2016).
The density profile for stars with $V_{lim}=16$ mag is shown in the left panel of Fig.~5.
The maximum distance from the center is taken one kernel halfwidth from the detector border
to avoid under-sampling (see previous section).
We estimated the kernel halfwidth
empirically by comparing of density profiles with the different kernel half-widths and selecting the smoothest curve, that closely follows the mean trend, defined by curves with much smaller kernel halfwidth values (Merritt \& Tremblay 1994, Seleznev 2016).
The density profile is shown with a solid line, while the border of 2$\sigma$--width confidence interval is shown
with two dotted lines. This interval was obtained by employing the smoothed bootstrap estimate method (Merritt \& Tremblay 1994,
Seleznev 2016). The visual estimate of the field stellar
density is indicated with a straight solid line, and was inferred as follows: if
the density profile gets flat within the covered region (it is
the case for NGC~4337), then the field density line is drawn in a way that the fluctuations around it
have equal areas.  Seleznev (2016) showed, that visual estimate of the field density correlates well with the
result of its determination with much more sophisticated method. As a consequence, the cluster
radius corresponds to the  intersection of the density profile with the field density level.
The associated uncertainty is computed using the intersections
of the confidence interval lines with the field density at the cluster radius location.
Seleznev (2016) showed that this estimate of cluster radius does not depend significantly
on the adopted kernel halfwidth value, when the adopted kernel halfwidth and the smaller ones are used.
The uncertainty of the field density estimate was evaluated as a half of the confidence interval width at the cluster radius point.

The cluster radius $R_c$ is considered here as the radius of the sphere around the cluster centre, where
the cluster differs from the field (Danilov, Matkin, \& Pylskaya 1985, Danilov \& Seleznev 1994).
This value doesn't coincide with tidal radius. The last one can be larger than $R_c$ (for example, if
the field density fluctuations prevent to detect outer low-density part of the cluster) or smaller
than $R_c$ (for example, if we detect the relatively dense part of the cluster tidal tails).
We don't fit radial surface density profiles with King model, because it doesn't reproduce well
the outer part of open clusters, which tends to underestimation of open cluster size (Seleznev 2016).

The use of radial density profiles implies an assumption of spherical symmetry. What effect this
assumption puts onto our results? Open cluster often are asymmetric (NGC 4337 is just the case).
The cluster radius obtained with the radial density profile corresponds in that case to the distance
from the cluster centre to the most remote point of the cluster boundary (if background density
fluctuations are not too large). Points of the density profile represent the azimuthally-averaged
values of density. Then the integral of the radial density profile will give the same result as
the integral of two-dimensional density distribution. It is important for the following treatment.

The estimates of cluster radius for different limiting magnitudes
are listed in the 2nd column of Table 4. The 3rd column contains the cluster radius in parsecs,
adopting the distance as in  Sect.~5.
The size of field under investigation is relatively small, and there
is a possibility that we are missing  the cluster halo. Therefore it is more conservative  to consider
these values as lower estimates of the cluster radius.
The surface
density profile derived using  2MASS (Fig.~5, right panel) lends further support to this scenario.
In fact, the cluster radius can be estimated in this case
as large as $R_c=10.0\pm0.8$ arcmin. A visual comparison with the density maps in Fig.~4 shows that this radius estimate
corresponds to the cluster maximum elongation in the North-East and the South-West directions. In the perpendicular
direction the cluster extent seems considerably smaller. Due to this reason we use the mean field stars surface
density $F_b$ (4th column of Table 4) as determined for triangles marked by white dotted line in the right panel
of Fig.4. It was obtained with the cumulative luminosity function for these regions.

Finally,  one can notice from Table 4 that brightest stars of
NGC 4337 (the RGB clump stars) are distributed in a smaller volume than fainter stars, which indicates that the cluster already
experienced some dynamical evolution and mass segregation.

\begin{table}
\tabcolsep 0.2truecm
\caption{Cluster parameters with different limiting magnitudes.}
\begin{tabular}{cccccc}
\hline
$V_{lim}$ &    $R_c$    &     $R_c$   &     $F_b$      &    $N_c$    &   $r_h$     \\
          &   arcmin    &       pc    & $arcmin^{-2}$  &             &   arcmin    \\
\hline
  14      & 4.6$\pm$0.3 & 2.9$\pm$0.2 & 0.30$\pm$0.09  &   33$\pm$19 & 2.0$\pm$1.4 \\
  15      & 6.7$\pm$0.7 & 4.3$\pm$0.4 & 0.66$\pm$0.16  &  117$\pm$53 & 2.3$\pm$1.1 \\
  16      & 6.2$\pm$0.5 & 4.0$\pm$0.3 & 1.68$\pm$0.26  &  198$\pm$72 & 2.4$\pm$0.9 \\
  17      & 7.0$\pm$0.2 & 4.5$\pm$0.1 & 3.0$\pm$0.4    &  392$\pm$122& 3.2$\pm$1.0 \\
  18      & 7.2$\pm$0.2 & 4.6$\pm$0.1 & 5.5$\pm$0.6    &  650$\pm$177& 3.7$\pm$1.0 \\
  19      & 6.7$\pm$0.4 & 4.3$\pm$0.3 & 11.5$\pm$0.8   &  758$\pm$218& 3.4$\pm$0.9 \\
  20      & 6.5$\pm$0.3 & 4.2$\pm$0.2 & 22.0$\pm$1.1   & 1128$\pm$280& 3.7$\pm$0.6 \\
  21      & 7.0$\pm$0.2 & 4.5$\pm$0.1 & 35.6$\pm$1.4   & 1931$\pm$395& 4.3$\pm$0.6 \\
\hline
\end{tabular}
\end{table}

We integrated the density profile  to estimate the cluster star number and the parameter $r_h$, namely
the radius of the circle around the cluster center that contains  half of the cluster stars. The integration was
performed using Simpson method with accuracy estimate, while the interpolation of density profiles was performed
by using  a spline function. Column 5 in Table~4 reports the cluster star number,  while the 6th column contains estimates
of $r_h$, and indicates that this increases at increasing  the limiting magnitude.
Usually, in the literature, $r_h$ is defined as the half light or mass radius, which not necessarily coincides with the
radius at which half of the stars are counted. Nevertheless $r_h$ is used for the virial mass estimate
of star cluster, while the accurate formula for virial mass contains the mean inverse star-to-star distance instead
of $r_h$. Therefore it is not very important in which way $r_h$ is determined.

The number of  cluster stars can be used to infer an estimate of the cluster mass, as illustrated in Table 5.
The first column contains the magnitude V and the second column lists the absolute  magnitude
$M_V$, calculated adopting the distance modulus derived in  Sect.~ 5. The third column contains the mass of star corresponding to this magnitude
taken from the table of isochrone for the age and the metallicity of NGC 4337 (Carraro et al. 2014b,
Bressan et al. 2012).
The fourth column contains the mean mass of the magnitude interval between this magnitude and
the magnitude from the previous raw (the mean mass value in the first row is for  brighter -RGB clump- cluster stars).
The fifth column lists the number of stars in this magnitude interval, and the sixth column contains
the mass estimate of cluster stars from the same magnitude interval.

\begin{table}
\tabcolsep 0.2truecm
\caption{Cluster mass estimate}
\begin{tabular}{cccccc}
\hline
  V   &       $M_V$     &      m         &     $<m_i>$    &    $N_i$    &  ${\mathfrak M}_i$  \\
      &        mag      &   $M_{\odot}$   &    $M_{\odot}$  &             &       $M_{\odot} $   \\
\hline
  14  &    1.28$\pm$0.02  &   1.90$\pm$0.00  &   1.95$\pm$0.00  &  33$\pm$19  &      64$\pm$37      \\
  15  &    2.28$\pm$0.02  &   1.66$\pm$0.01  &   1.78$\pm$0.00  &  84$\pm$56  &     150$\pm$100     \\
  16  &    3.28$\pm$0.02  &   1.39$\pm$0.01  &   1.53$\pm$0.01  &  81$\pm$89  &     124$\pm$136     \\
  17  &    4.28$\pm$0.02  &   1.17$\pm$0.01  &   1.28$\pm$0.01  & 194$\pm$142 &     248$\pm$182     \\
  18  &    5.28$\pm$0.02  &   1.00$\pm$0.00  &   1.09$\pm$0.01  & 258$\pm$215 &     281$\pm$234     \\
  19  &    6.28$\pm$0.02  &   0.86$\pm$0.01  &   0.93$\pm$0.01  & 108$\pm$281 &     100$\pm$261     \\
  20  &    7.28$\pm$0.02  &   0.74$\pm$0.00  &   0.80$\pm$0.01  & 370$\pm$355 &     296$\pm$284     \\
  21  &    8.28$\pm$0.02  &   0.63$\pm$0.00  &   0.69$\pm$0.00  & 803$\pm$484 &     554$\pm$334     \\
 \hline
\end{tabular}
\end{table}

We anticipate here that the cluster luminosity function (see below) shows a sharp decrease beyond $V\simeq 21$ mag,
despite of completeness correction. This cut-off is most probably produced by large uncertainty in the completeness
factor at magnitudes close to the investigation limit and by selection effects. Because of this, the cluster
mass was estimated only down to $V=21$ mag, and results to be
${\mathfrak M}=1820\pm 620\; M_{\odot}$. This is clearly intended as a lower estimate of the cluster actual mass,
because the mass of 0.63 $M_{\odot}$ is far from the lower stellar mass limit, and the cluster radius
values in Table 4 are the lower estimates (see above).

\begin{figure*}
\includegraphics[height=8.2truecm]{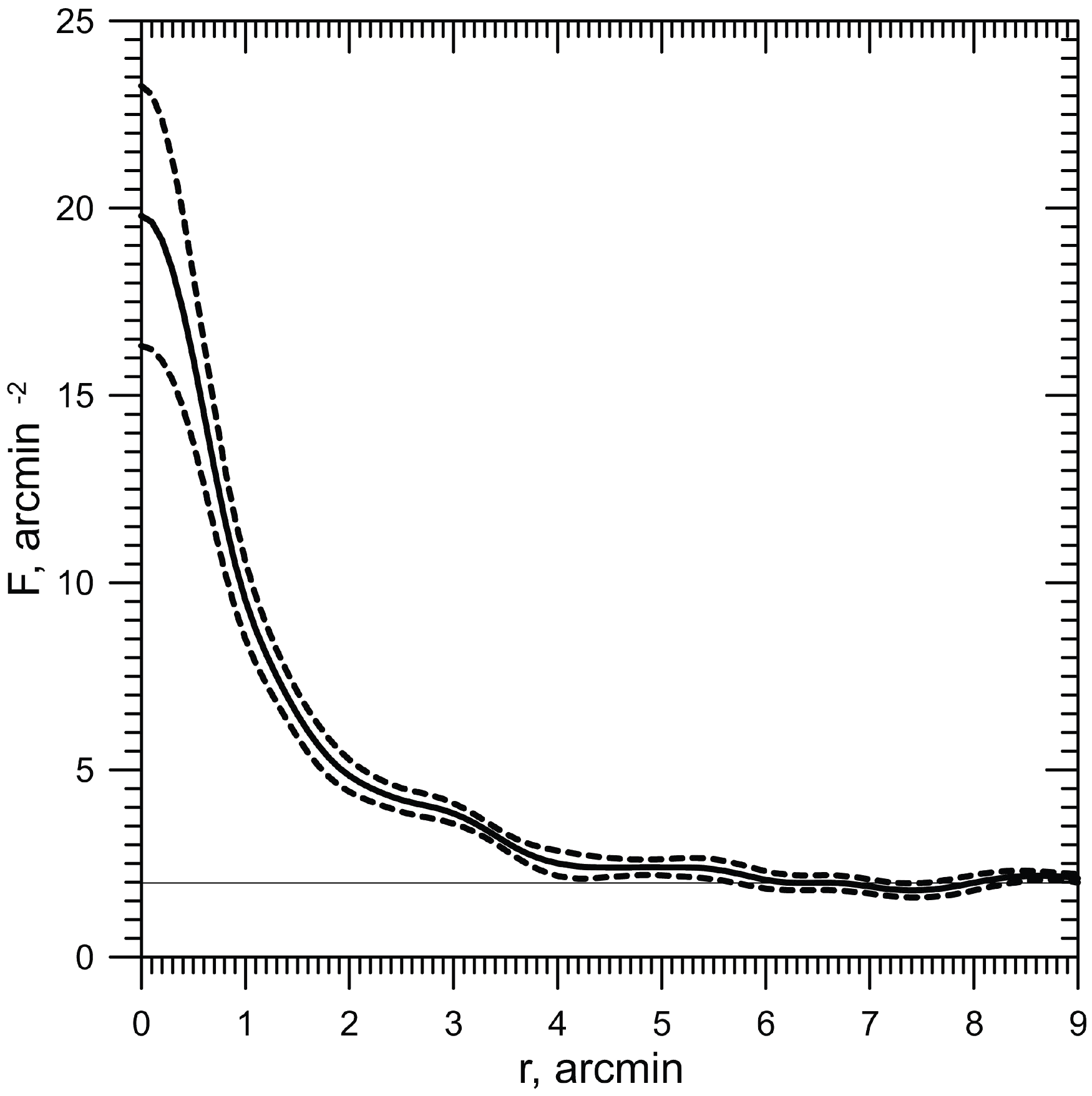}
\includegraphics[height=8truecm]{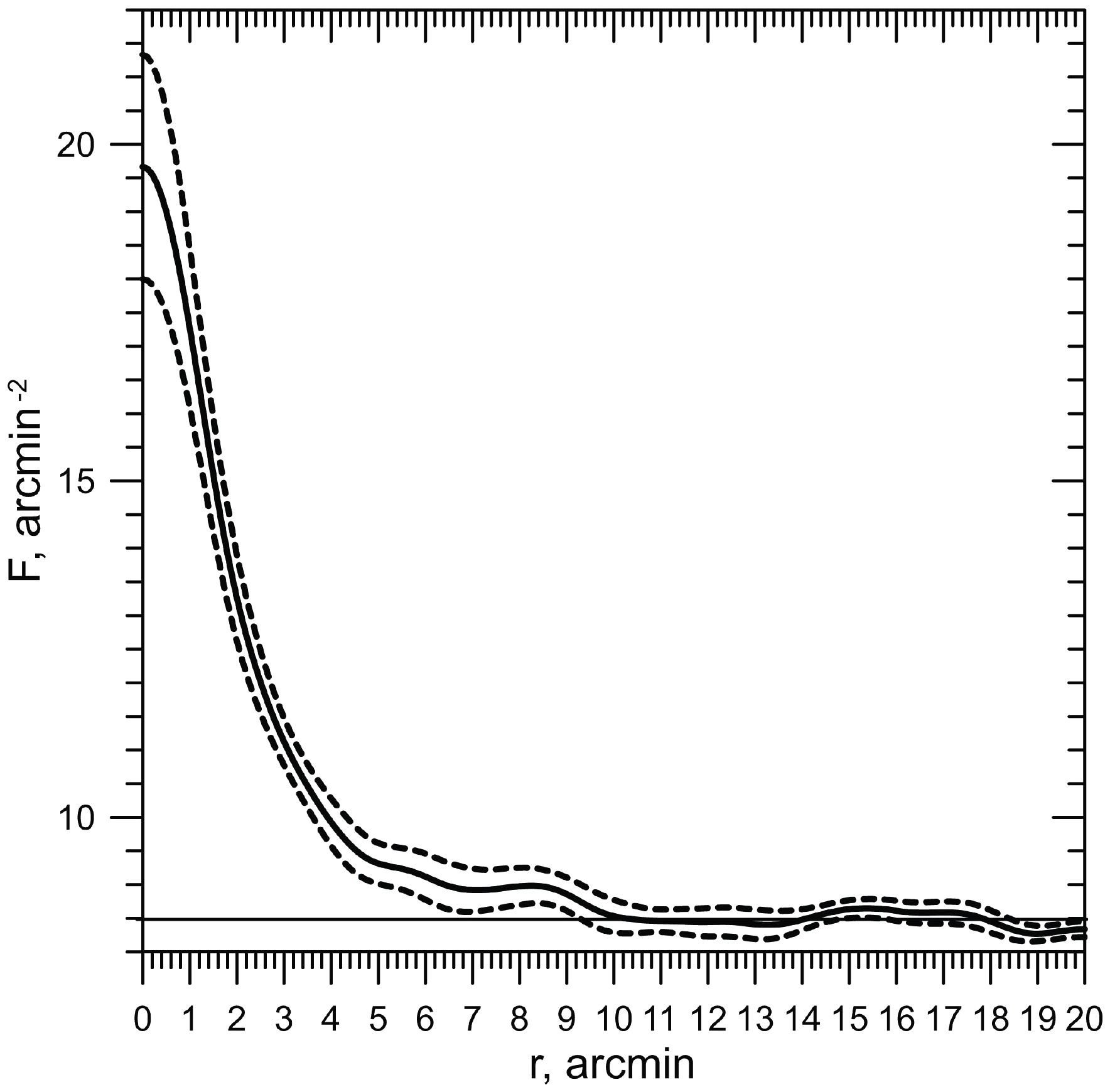}
\caption{{\bf Left panel}: the radial surface density profile of the cluster for $V_{lim}=16$ mag. {\bf Right panel}:  the radial surface density profile of the cluster with the stars of $J_{lim}=16$ mag, obtained with the data
of 2MASS. The solid line shows density profile obtained by kernel estimator with the
kernel half-width of 3 arcmin for left panel and 2 arcmin for right panel. The dashed lines show the
2$\sigma$--width confidence interval. The straight solid
line shows the value of the field surface density determined with the density profiles.
}
\end{figure*}

\section{The cluster luminosity function}
\label{clf}
\noindent
The same kernel estimator method was employed  to construct the cluster luminosity function (LF, see, e.g.,  Prisinzano et al. 2001).
However, at odds with  Prisinzano et al. (2001), we adopt here a fixed (not adaptive) kernel,
which is more effective in the presence of the rich stellar field around NGC~4337, since
it gives the same result with less computational effort.

We estimated the LF in the cluster area (central rectangle,
marked by white dotted line in Fig.~4, right panel), and in the comparison region (two triangles in the corners between the white
dotted lines and the boundary, marked by white solid line). Due to elongated shape of NGC 4337 we expect these two
regions do not contain cluster stars and effectively represent the
stellar field in the cluster vicinity. It is worth noticing that the area of the cluster region is
equal to the area of comparison fields, and that we took into account the results of incompleteness analysis.

The LF of the cluster stars was then derived as the difference between the LFs of the cluster region (Fig.~6, left panel) and
the comparison region (Fig.~6, middle panel). This LF is shown in right panel of Fig.~6 with a thick solid line, while thin solid lines indicates
the 2$\sigma$--width confidence interval.
The kernel halfwidth was taken to be 0.5 mag, it was chosen by the same consideration, as in the case of the density profiles (see above in Section 6). The kernel estimate is a continuous and differentiable function, and it is very important for further mass function evaluation (Seleznev 2016).

One can notice  that below  $V=21$ mag the LF of cluster stars has a sharp decrease, despite of completeness correction.
The most probable explanation is that completeness factors have large uncertainties near the investigation limit,
and we reached the completeness limit of our photometric dataset (see Table~2).
In the following we therefore restrict ourself to stars brighter than $V=21$ mag.
The CMD (Fig.~3) shows
that the brightest cluster stars have magnitudes in the interval $V\in[13.5;14.3]$ mag, and therefore
 $V=13$ mag seems  quite a reasonable magnitude to start the LF computation with. Negative values of LF are not impossible,
 because this was obtained as a
difference between two distributions. The negative LF regions (near $V=12.5$ mag and $V=22$ mag) are in any case outside
the region of our interest. Finally, we draw the attention to a LF minimum near $V=16$ mag and $V=18$ mag.
We run a Kolmogorov-Smirnov (KS) test that does not show that these minima are statistically significant, but the chi-square test shows noticeable difference with a  p-value of about 0.17. The comparison of LF (and its histogram in the case of chi-square test) was carried out with the curves (and histograms), where the minima were replaced by the graduate positive slope. In the case of KS test a cumulative LF was constructed.

It is crucial to underline that this LF is normalised to the cluster star number inside the white-dotted-line rectangle
(see right panel in Fig.~4). In order to obtain a normalisation to the whole  cluster we need to derive the ratio of the cluster star
number inside the rectangle to total cluster star number. The cluster star number inside the rectangle was
derived as the double integral of the surface density profile over the rectangle, and the integration was made with the
method of rectangles in two dimensions. Then the field star number was subtracted from the result of
this integration. The total star number and the surface density of field stars were both taken from Table~4.
The normalisation ratio was then determined for limiting magnitude $V_{lim}=21$ mag to be $0.55\pm0.24$.

\begin{figure*}
\includegraphics[height=5.5truecm]{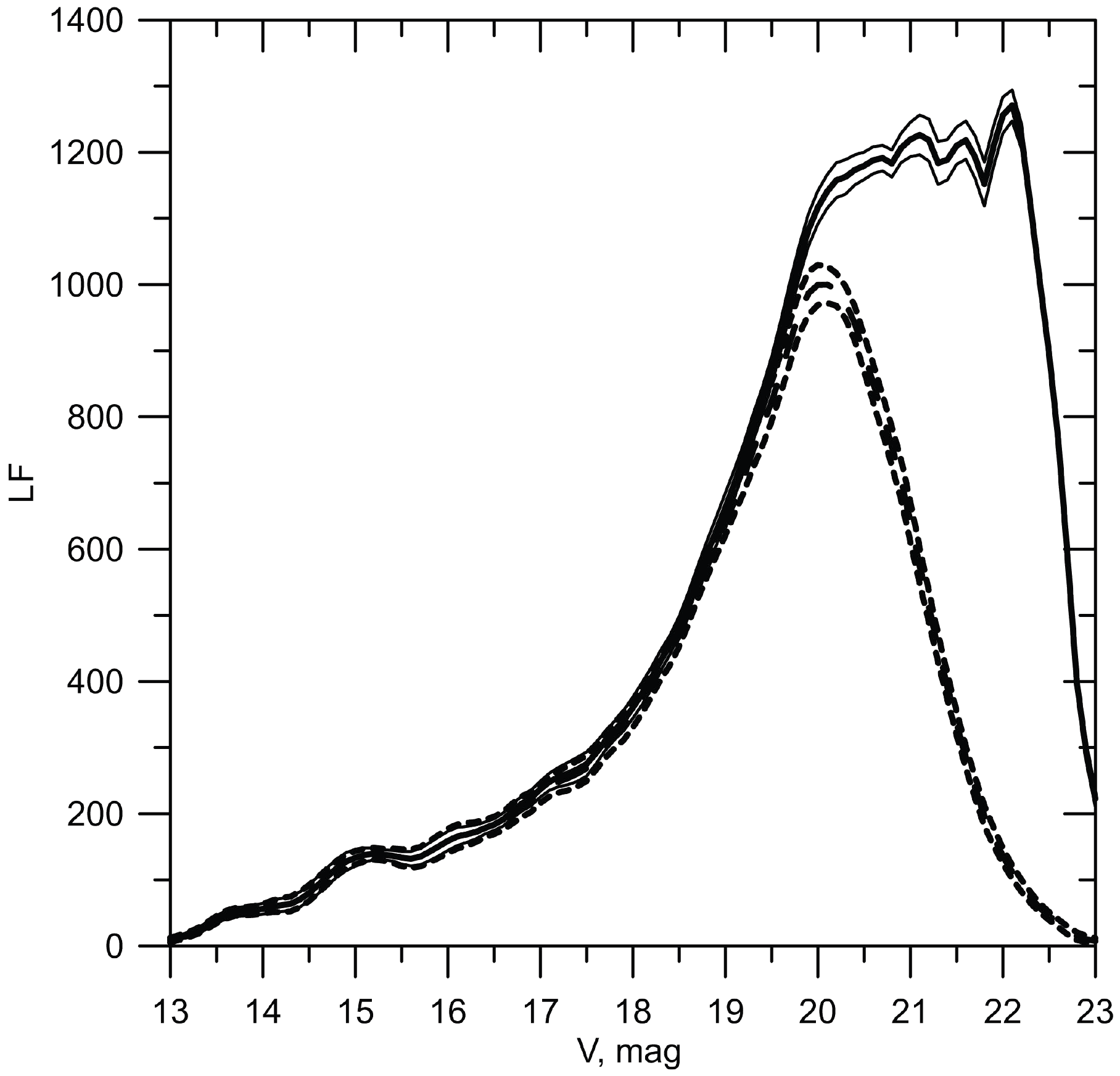}
\includegraphics[height=5.5truecm]{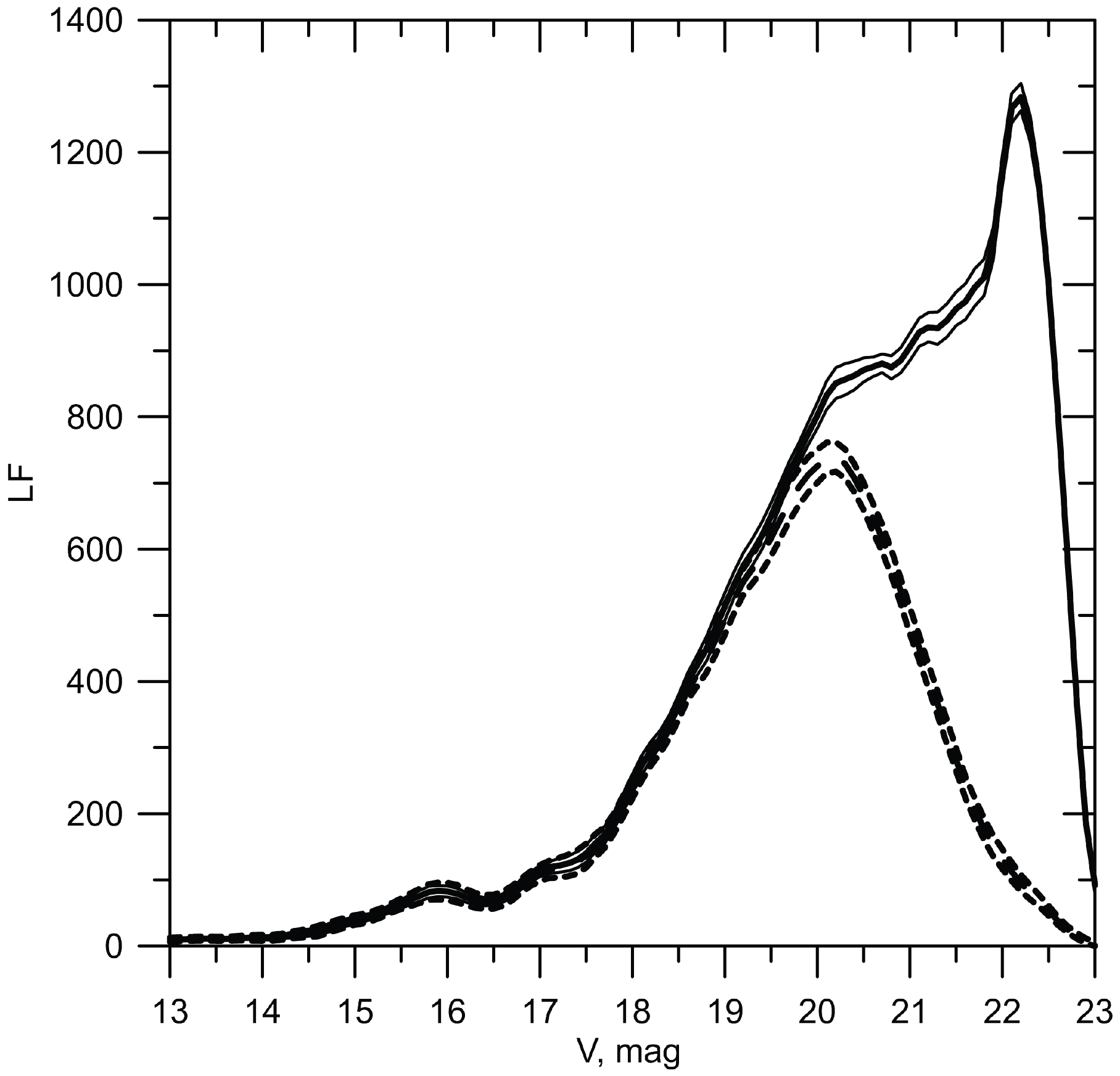}
\includegraphics[height=5.5truecm]{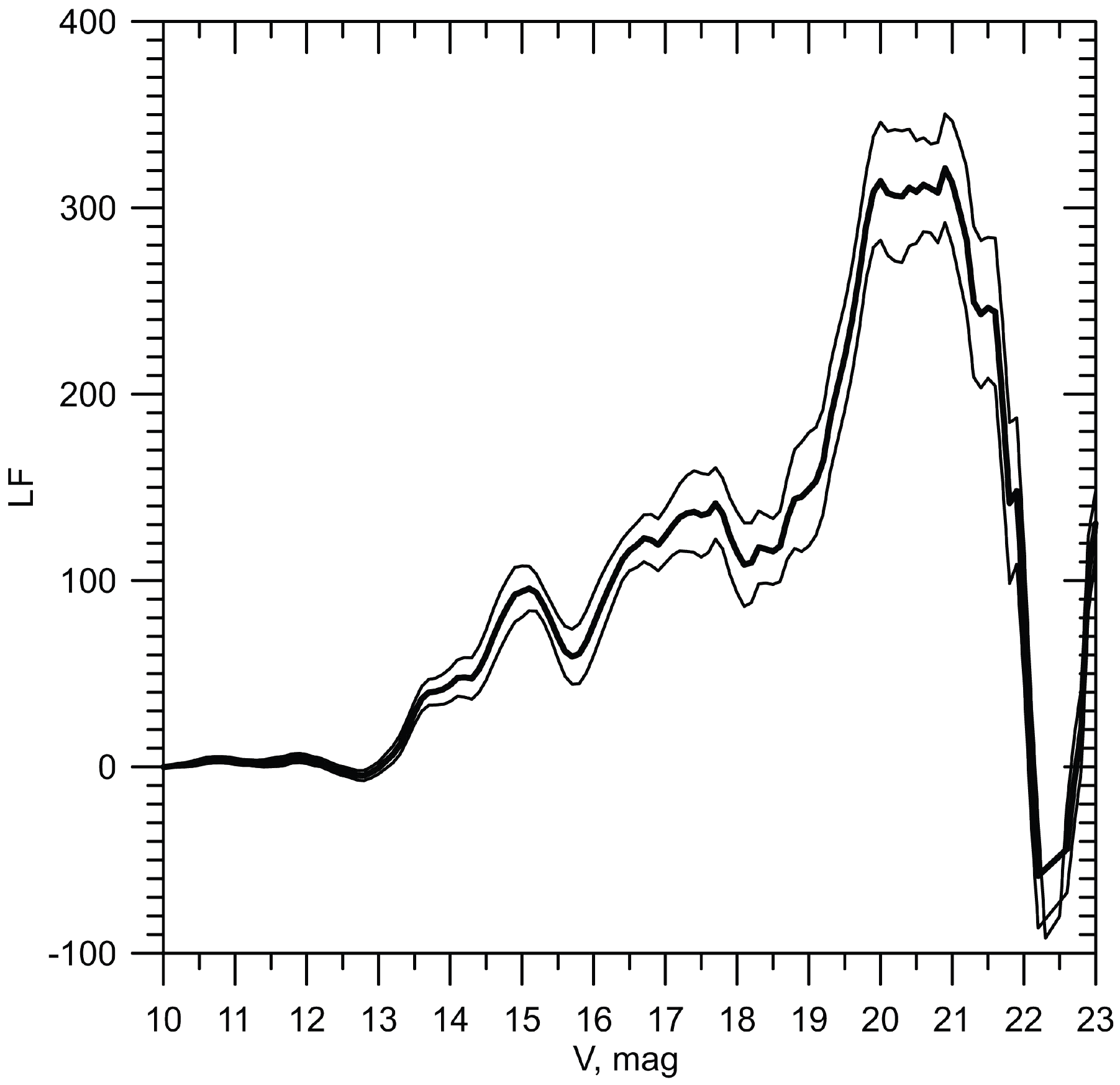}
\caption{The luminosity function of NGC~4337. {\bf Left panel}: The thick solid line shows the LF for cluster region
 (white-dotted square in the Fig.4, right panel), corrected for incompleteness. The LF was obtained by kernel estimator
  with the kernel halfwidth of 0.5 magnitude. Thin solid lines show the 2$\sigma$--width confidence interval.
  Dashed lines show uncorrected LF for this region with its confidence interval. {\bf Middle panel}: The same as in the
  left panel, but for field region (triangles separated by white dotted lines in the Fig.4, right panel).
  {\bf Right panel}: The LF for cluster stars, the result of subtraction of the field LF (thick solid line in the
  middle panel) from the cluster region LF (thick solid line in the left panel), with its 2$\sigma$ confidence interval.}
\end{figure*}

\section{The cluster mass function}
\label{clusmf}
Let consider the cluster mass function (MF) $\psi(m)$ as:

\begin{equation}
\label{MF}
\psi(m)=\frac{dn}{dm}\; \mbox{,}  \qquad \int\limits_{m_1}^{m_2} \psi(m)dm=N \; \mbox{,}
\end{equation}

\noindent
where $m$ is the mass of a star and N is the number of cluster stars in the mass range of $[m_1;m_2]$.
The cluster mass ${\mathfrak M}$ for this mass range is determined then as:

\begin{equation}
\label{mass}
\int\limits_{m_1}^{m_2} \psi(m)mdm={\mathfrak M} \; \mbox{.}
\end{equation}

\noindent
Using the same terminology, the luminosity function can be written as

\begin{equation}
\label{LF}
\varphi(V)=\frac{dn}{dV}\; \mbox{,}  \qquad \int\limits_{V_1}^{V_2} \varphi(V)dV=N \; \mbox{.}
\end{equation}

\noindent
If the mass range corresponds to the magnitude range, the cluster star number N will be the same.

\noindent
Let then $m=m(V)$ be the stellar mass-luminosity relation. In this case, we have:

\begin{equation}
\label{LF-MF}
dm=\frac{dm}{dV}\cdot dV\equiv m'_V \cdot dV\; \mbox{, and}
\end{equation}
\begin{equation}
\psi(m)=\frac{dn}{dm}=\frac{dn}{|m'_V|\cdot dV}=\frac{\varphi(V)}{|m'_V|} \; \mbox{.}
\end{equation}

\noindent
The LF was converted into MF in the magnitude range $V\in [14.5;21]$ mag, which corresponds to the absolute magnitude
range $M_V\in [1.78;8.28]$ mag. This ensure we avoid any ambiguity in the region of RGB clump stars and selection effects
in the region $V>21$ mag.
The relation  $m=m(V)$ was taken from Padova suite of models (Bressan et al. 2012), and
was approximated by a spline function, together with its first derivative. The cluster mass
function is shown in Fig.~7 (left panel) in the linear form and in Fig.~7 (right panel) in the logarithmic form.

A least squares regression over the logarithmic MF yields a MF slope of -2.68$\pm$0.08 (in this scale
the standard Salpeter slope is -2.35).
In order to take into account the
confidence interval for MF, numerical experiments have been performed. For every $m$ point the value of MF
was randomly taken from the interval $[MF-3\sigma;MF+3\sigma]$ according to a  gaussian distribution
with its centre at MF value. Then a linear regression was performed with an even series of MF values. The error
of the argument (the mass) was not taken into account. Twenty experiments produced  a mean slope -2.72$\pm$0.08,
virtually identical to the values obtained by the simple linear regression.

A lower estimate of the cluster mass was then derived by integrating the MF over the whole mass range. This yielded a
mass
of $976\pm 135 \;M_{\odot}$ for stars inside the white-dotted-line rectangle in Fig.4 (right panel), and in the magnitude range
$V\in [14.5;21]$ mag.
Assuming the same MF for the entire  cluster, the lower estimate of total cluster mass in the same
magnitude range would be $1775\pm 812 \;M_{\odot}$ (the normalization ratio from the previous section is applied).
If in addition we account for the stars with $V\leq14.5$ mag from the surface
density profile for stars with $V_{lim}=14.5$ mag, this estimate would become  $1880\pm 820 \;M_{\odot}$.
Notice that the possible (small) bias toward high luminosities (masses) in our LF sample, caused by dynamical mass segregation,
implies that the above mass estimate should be considered as an upper limit.

The reader can notice that this mass estimate does not differ significantly from the estimate obtained by surface density profiles in the previous section, which is ${\mathfrak M}=1820\pm 620\; M_{\odot}$ (see above).
These both estimates are anyway still a lower limit for the mass estimate, because of the unknown low-mass end
of the stellar mass distribution, and the un-accounted unresolved binaries and probable remnants of massive stars.

\begin{figure*}
\includegraphics[height=8truecm]{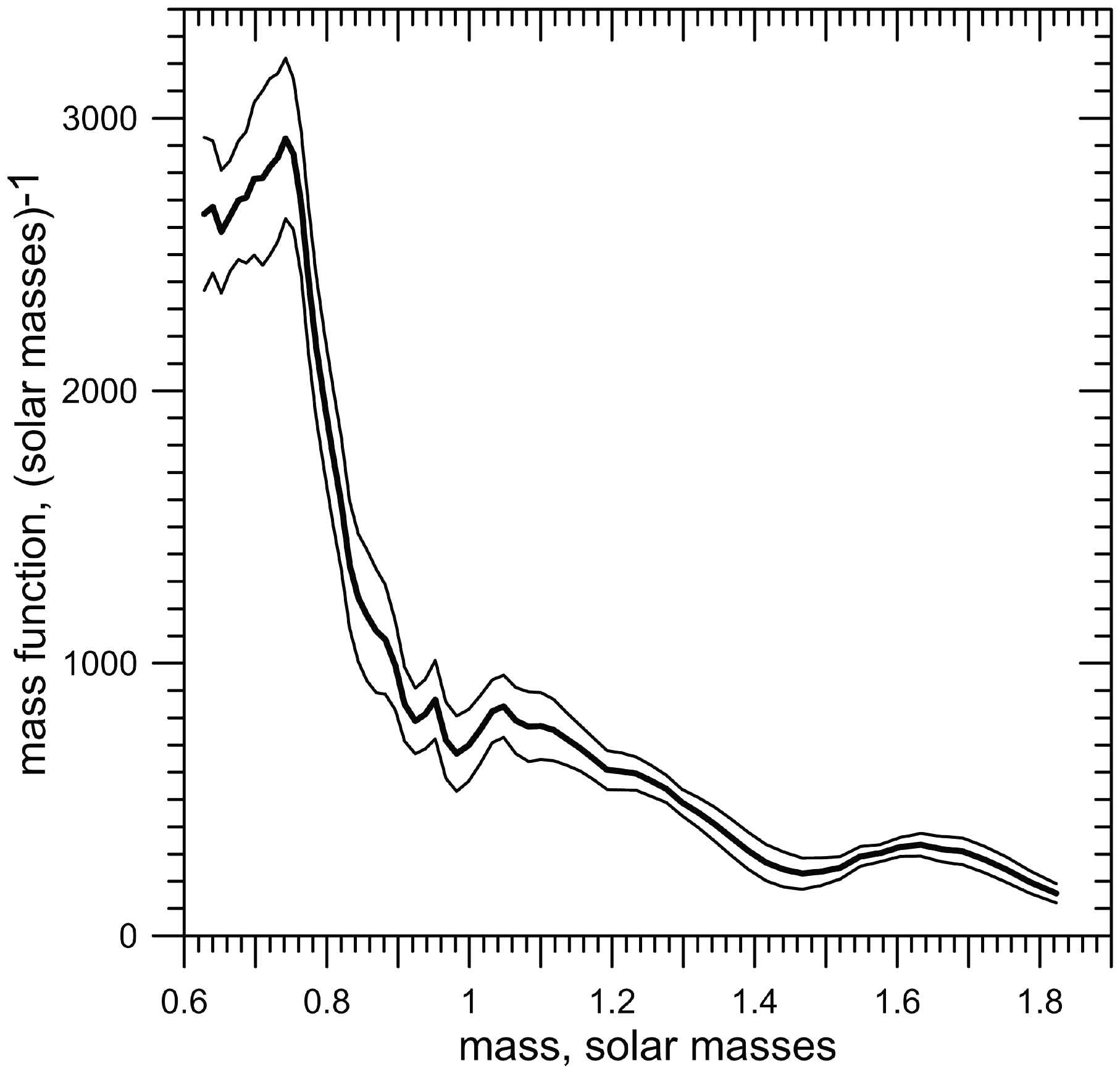}
\includegraphics[height=8truecm]{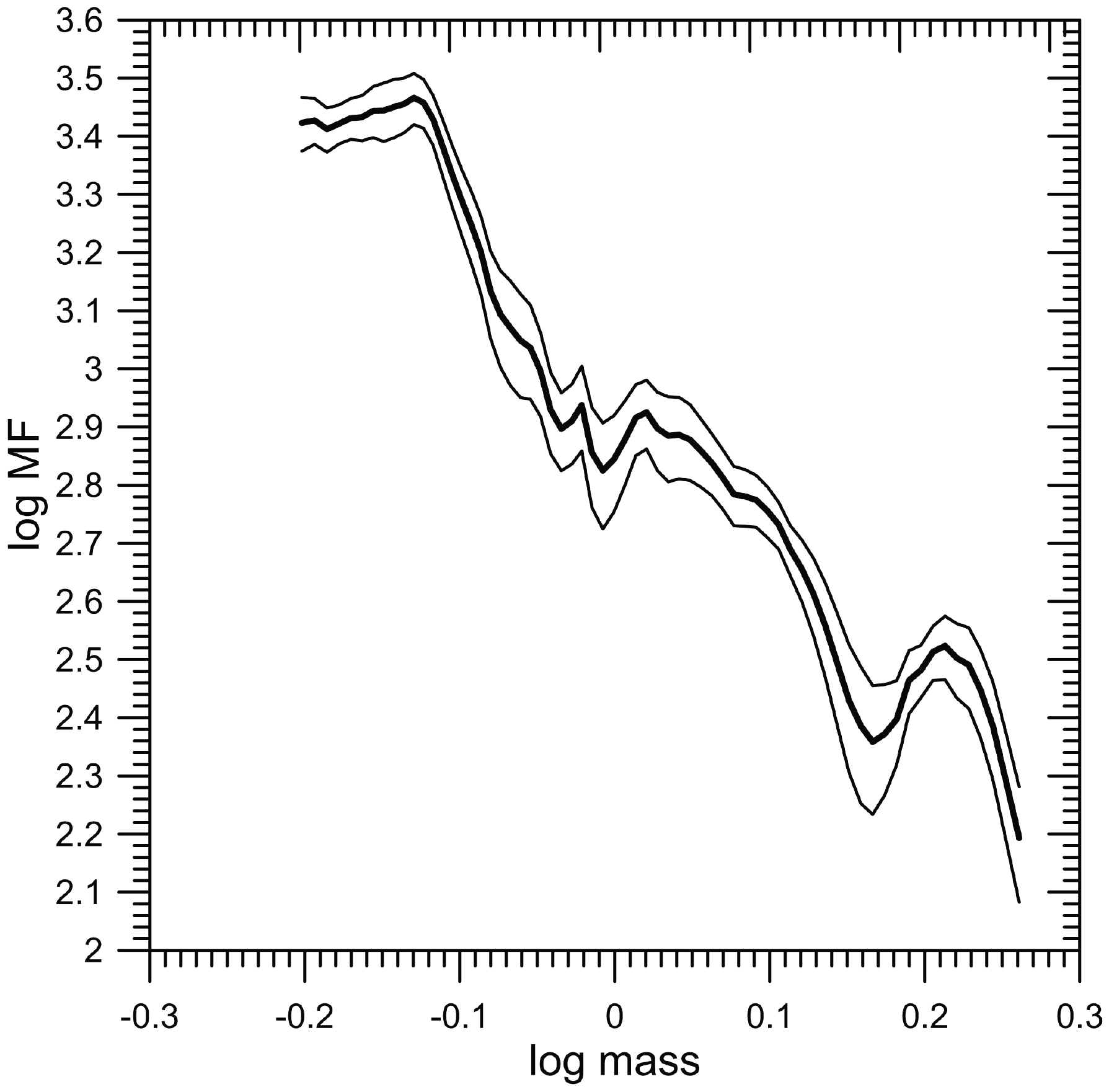}
\caption{{\bf Left panel} : The mass function of the cluster. The thick solid line shows LF, thin solid lines show the
2$\sigma$--width confidence interval.
{\bf Right panel}:  The mass function of the cluster in logarithmic form. Symbols are the same as in the left panel.}
\end{figure*}

\section{Cluster dynamical mass}
\label{dynamical}
\subsection{Isolated cluster}

The most common, but rough, way to obtain a cluster dynamical mass estimate is by mean of the computation of the 1D (radial) velocity dispersion for a set of $n$ stars for which the radial velocity is available.
This gives

\begin{equation}
\sigma_{r}^2=\frac{\sum_{i=1}^n \left(v_{r,i} - \langle  v_r \rangle \right)^2}{n},
\end{equation}

\noindent
where $\langle v_r \rangle$ is the mean radial velocity, which corresponds to the cluster sampling systemic velocity.
In the case of NGC 4337, data in Table~3 give $\langle v_r\rangle = -17.78\pm 1.00$  kms$^{-1}$ and $\sigma_r= 1.62\pm0.30$ kms$^{-1}$
A virial mass is obtained as

\begin{equation}
M=\frac{R\sigma^2}{\alpha G},
\label{vm1}
\end{equation}

\noindent
where $\sigma$ is the 3D velocity dispersion which, in the assumption of isotropic velocity distribution, is $\sigma^2=3\sigma_r^2$, $R$ is the cluster radius, $\alpha$ is a geometric factor such that $\Omega = -\alpha GM^2/R$ with $\Omega$ the cluster gravitational energy.
Applying eq.~7 to our data we get $5.26\times 10^3 \alpha^{-1} \leq M (\rm{M}_\odot) \leq 8.34\times 10^3 \alpha^{-1}$, adopting the minimum (2.9 pc) and maximum (4.5 pc) value for $R$ in Table 4.

The choice of $\alpha =3/5$ (homogeneous cluster) gives $8.77\times 10^3 \leq M(\rm{M}_\odot) \leq 1.93\times 10^4$, while a factor 2 larger $\alpha$ (much more compact cluster) leads to  $4.39\times 10^3 \leq M(\rm{M}_\odot) \leq 9.65\times 10^3$, all values significantly larger than the estimates based on the radial density profile in Sect. \ref{raddenprof}, $1820$ M$_\odot$, or on the cluster MF in Sect. \ref{clusmf}, $1880$ M$_\odot$.

Actually, the ratio between the dynamical mass estimate and the density profile or LF based estimates is in the range $2.34 \div 10.63$, which indicates either a very large quantity of invisible mass or that the cluster is not virialised.

What we said above suggests as important to deepen the topic of dynamical mass estimates.
At this regard, we developed a simple formula based on the knowledge of a limited sample of $n$ angular positions and radial velocities, as follows.
The 1D kinetic energy in the cluster rest frame is straightforwardly obtained as

\begin{equation}
K_1=\frac{1}{2}\sum_{i=1}^n m_i (v_{r,i}-\langle v_r\rangle)^2
\end{equation}

\noindent
while the gravitational potential energy is

\begin{equation}
\Omega = -\sum_{j>i}G\frac{m_im_j}{r_{ij}},
\end{equation}

\noindent
where $r_{ij}\equiv |{\bf r}_i-{\bf r}_j|$ is the distance between the $i-th$ and $j-th$ star in the system, whose position vector are ${\bf r}_i$ and ${\bf r}_j$ with $r_i$ and $r_j$ their moduli. Given the angular coordinates (right ascension and declination, hereafter $\theta$ and $\phi$) of the stars, we have

\begin{equation}
r_{ij}=\sqrt{r_i^2+r_j^2-2r_ir_j\left[\cos\phi_i
\cos\phi_j+\sin\phi_i\sin\phi_j\cos(\theta_i-\theta_j)\right]}
.
\label{dis}
\end{equation}

\noindent
The assumption that the generic star distance to the reference frame origin, $r_i$, is the same (the cluster distance, $d$), leads to

\begin{equation}
r_{ij}=\sqrt{2}d\sqrt{1-\left[\cos\phi_i
\cos\phi_j+\sin\phi_i\sin\phi_j\cos(\theta_i-\theta_j)\right]}
.
\label{prodis}
\end{equation}

\noindent
The pair distances computed with Eq. \ref{prodis} are under-estimated, so a better approximation, once the cluster size $R$ is known, is to insert in Eq.\ref{dis} $r_i$ and $r_j$ randomly sampled within the cluster radius, via the simple linear expression

\begin{equation}
r_i=d-R+2Rt,
\end{equation}

\noindent
where $t$ is a random variable in the $[0,1]$ interval.

We computed the cluster potential energy, that we call $\Omega_3$, using the latter approximation for the pair distances while for the kinetic energy we adopted the natural assumption of velocity isotropy that leads to a 3D kinetic energy for the $n$ stars sample which is simply $K_3=3K_1$.

Note that although we took into account, in an approximate way, the 3D structure of the cluster the virial ratio obtained $Q_3= 2K_3/|\Omega_3|$ is expected to be an overestimate of the actual virial ratio even in absence of non luminous matter, because the sample kinetic energy is $O(n/N)$ the total cluster kinetic energy while the gravitational energy scales quadratically, so that $\Omega^{(n)} \propto (n/N)^2 \Omega^{(N)}$
(N is here the total number of the cluster stars).

Given this, a better estimate than eq. \ref{vm1} for a virial mass is that of the mass  which {\it closes} gravitationally the cluster:

\begin{equation}
M_3=Q_3M_*,
\end{equation}

\noindent
where $M_*=68.52$ M$_\odot$ is the mass of the 45 member stars for which radial velocities and angular coordinates are given in Table \ref{tab4}. Note that the value of M$_*$ is about 3.7\% of the masses evaluated by the surface density profile (Sect. \ref{raddenprof}) and via the luminosity function in Sect.~8.
Adopting error propagation formula giving the safest error estimate, we get $M_3 = 11400 \pm 4550$ M$_{\odot}$. This error does not include the error propagation due to uncertainty in the star radial velocities, distances and angular coordinates.

An important question to answer is: can the low mass, faint stars of the cluster and the high mass remnants account for the huge difference between this virial mass and the masses estimated via the density profile and the MF of previous Sections?
The answer to this question comes from the evaluation of the quantity of mass accounted by the power law mass function of Sect. \ref{clusmf}, $\psi(m) \propto m^{-s}$, where $s$ was taken at their nominal values of $-2.68$ and $-2.72$, cutting the MF at the lowest value $m_{min}=0.1$ M$_{\odot}$ (roughly the brown dwarf limit) and $m_{max}=25$ M$_{\odot}$. Note that the precise assumption on $m_{max}$ is not relevant because of the rapid decrease of the MF for large mass values.
The magnitude interval of visible stars is $[1.78;8.28]$ mag which transforms into the $[0.63;1.78]$ M$_{\odot}$ mass interval.
In Table~6 we report the values of the mass contributed by the cluster stars out of this mass interval (low mass, index $l$, and high mass, index $h$) and the mass given by visible stars. Values in this Table essentially show how dark stars (in the low and high mass tail of the MF) can marginally provide the undetected mass to give a virialised cluster without claiming for dark matter.
At this regard, we additionally note that the fraction of mass contributed by high and low mass stars reported in Table~6 is an upper limit, because we did not account for the mass loss from the ZAMS to the remnants and for stars escaping the clusters over its life-time.


\begin{table}
\label{dynmas}
\tabcolsep 0.3truecm
\caption{MF exponent, low mass to visible stars number ratio, high mass to visible stars number ratio, low mass to visible stars mass ratio, high mass to visible stars mass ratio, total invisible (low + high mass) to visible stars mass ratio}
\begin{tabular}{cccccc}
\hline
 s & $N_l/N_v$ & $N_h/N_v$ & $M_l/M_v$ & $M_h/M_v$ & $M_i/M_v$\\
\hline
-2.68  & 0.341 & 30.5 & 1.24 & 5.61 & 6.85\\
-2.72  & 0.305 & 36.6 & 1.07 & 6.74 & 7.81\\
\hline
\end{tabular}
\end{table}

\subsection{The contribution of the Milky Way gravitational field}

Danilov \& Loktin (2015) proposed a formula for a dynamical evaluation of a star cluster mass accounting for the Galactic gravitational field and for non-stationarity of the cluster:

\begin{equation}
\label{mass_dan}
M_d=\frac{2\bar RR_u\left[2\sigma^2-\displaystyle{\frac{(\alpha_1+\alpha_3)<r^2>}{3}}\right]}{G(\bar R+R_u)}  \; \mbox{,}
\end{equation}

\noindent
where $R_u=<1/r_i>^{-1}$ is the mean inverse star distance to the cluster centre,
$\bar R=<1/r_{ij}>^{-1}$ is the mean inverse star-to-star distance,
$<r^2>$ is the mean square of the star distance to the cluster centre, $\alpha_1$ and
$\alpha_3$ are the field constants (Chandrasekhar 1942)
characterising the Galactic
potential, $\Phi(R,z)$ in Galacto-centric cylindrical coordinates, in the vicinity of a star cluster:

\begin{equation}
\label{alpha1}
\alpha_1=\left(\frac{1}{R}\frac{\partial\Phi}{\partial R}-\frac{\partial^2\Phi}{\partial^2 R}\right)_{R_{cl}}=4A(B-A)<0  \; \mbox{,}
\end{equation}

where $A$ and $B$ are the Oort's constants,
and

\begin{equation}
\label{alpha3}
\alpha_3=-\left(\frac{\partial^2\Phi}{\partial^2 z}\right)_{z_{cl}}>0  \; \mbox{.}
\end{equation}

$R_{cl}$ and $z_{cl}$ are the cluster center of mass cylindrical coordinates.
The values of $\alpha_1$ and $\alpha_3$ were calculated adopting the Galactic potential model of
Kutuzov \& Osipkov (1980). Arguments in favour of this model are listed in Seleznev (2016).

The error of the radial velocity dispersion was estimated by the formula
$D\sigma^2\approx 2\sigma^4/N$ from Cramer (1946).

To estimate values of $\bar R$, $R_u$ and $<r^2>$ we obtain the spatial
distribution of stars around the cluster center by Monte Carlo sampling of the spatial density profile $f(r)$ as obtained by de-projecting the observed surface density profile $F(r)$ of the cluster (von Zeipel \& Lindgren 1921). Our technique requires a numerical differentiation of $F(r)$ which is not a problem because we adopt a kernel estimate of $F(r)$ which is a differentiable function.

\begin{equation}
\label{spadens}
f(r)=\frac{1}{\pi}\int \limits_0^{\sqrt{R_c^2-r^2}}S(\sqrt{r^2+x^2})dz \; \mbox{,}
\end{equation}

\noindent where

\begin{equation}
\label{spadens2}
S(r)=-\frac{1}{r}\frac{dF(r)}{dr} \; \mbox{,}
\end{equation}

Twenty different Monte Carlo samples were built, in order to estimate the scatter in the estimates.
For the spatial density profile corresponding to $V_{lim}=20$ mag following
estimates were obtained: $\bar R=2.71\pm0.14$ pc, $R_u=1.84\pm0.16$ pc, and
$<r^2>=7.20\pm0.71$ ${\rm pc^2}$ (the spatial density profile for $V_{lim}=20$ mag was choosing
because for this limiting magnitude the surface density profile is steadily decreasing,
it is of critical importance for de-projecting).

With these values of the mean stellar distribution characteristics the following
estimates of the cluster masses were obtained: $M_{vir}=10100\pm2200 \; M_{\odot}$
and $M_d=8200\pm2500 \; M_{\odot}$, where

\begin{equation}
\label{virial mass}
M_{vir}=\frac{2\sigma^2\bar R}{G} \; \mbox{.}
\end{equation}

There are two possible explanation for the large difference between
the dynamical and virial mass estimates from one side and the star-count mass estimates from another side.
One possibility is that star counts did not reveal the vast cluster corona (due to relatively small field, or due to large fluctuation of the field stellar density in the case
of star counts with 2MASS catalogue) so that the mass estimates obtained by star counts and the mass function are underestimated.
Remind, that mass estimates obtained with density profiles or mass function are lower ones because of the unknown low-mass end
of the stellar mass distribution, and the un-accounted unresolved binaries and probable remnants of massive stars.

Another possibility is that the velocity dispersion, upon which dynamical estimates are based, is overestimated. One reason for that
could be binarity, because the binary revolution orbital motion tends to enlarge velocity dispersion.

We can use Eq. 14 to estimate what value of velocity dispersion corresponds to the cluster mass estimates, obtained with the surface density profiles and with the mass function. The mass estimate $1880\pm820\; M_{\odot}$ corresponds to the total dispersion of
$1.39\pm0.38$ km/s and the radial velocity dispersion of $0.80\pm0.22$ km/s, assuming $\sigma_r=\sigma/\sqrt{3})$.
The mass estimate $1820\pm620\; M_{\odot}$ corresponds to the total dispersion of $1.37\pm0.29$ km/s and the radial velocity dispersion of $0.79\pm0.17$ km/s.

\section{Summary and conclusions}
In this work we exploited photometric and spectroscopic material to derive the present-day mass of the Galactic star cluster NGC~4337.
The star-count mass was derived both from the cluster density profile ($1820\pm620\; M_{\odot}$) and from the cluster luminosity function
($1880\pm820\; M_{\odot}$). The two estimates generally overlap. The dynamical mass, estimated in the assumption of virial equilibrium,
results to be significantly larger ($M_3 = 11400 \pm 4550$ M$_{\odot}$).
Considering the Galactic gravitational field and non-stationarity of the cluster results
with smaller estimate of $M_d=8200\pm2500 \; M_{\odot}$. Anyway, this smaller estimate at least 4.4 times exceeds the star-count mass. 
The possible contribution of invisible low mass stars and high mass
star remnants increase the cluster mass considerably (7.85 or 8.81 times depending on the adopted slope of the
cluster mass function, see Table 6). The luminous and dynamical mass could become comparable, if one takes into account this contribution
and the mass loss from the ZAMS to the remnants and stars escaping the cluster over its life-time.
Another possible sources of the discrepancy between star-count and dynamical mass estimates could be undetected vast cluster corona
(due to relatively small field, or due to large fluctuation of the field stellar density) or overestimated velocity dispersion value
(for example, due to inclusion of unresolved binary stars into the sample).

It is interesting at this point to try to reconstruct whether NGC~4337 suffered from significant mass loss, and
infer an estimate of its initial mass. This would help us to understand the reasons why it appears less dynamically evolved than its twin NGC~752.

We estimated the initial mass of the cluster by using three different measurements of the current mass of NGC\,4337.
We make use of an approximate method accounting for the mass loss due to stellar evolution, the Galactic tidal field, and encounters with giant molecular clouds and spiral arms.
We follow Dalessandro et al. (2015), who carried the same exercise for the old metal rich cluster NGC\,6791 and we refer the reader to their work for the detailed description of the procedure which is based on the theoretical study by Lamers et al. (2005).
To derive an estimate of the cluster initial mass, several constants characterising the cluster and the environment need to be adopted.
To evaluate the equation (1) of Dalessandro et al. (2015), we use $t=1.6\pm0.1$\,Gyr for the age of the cluster, and the three mass estimates derived above:
 $m_1=1820\pm620 M_{\odot}$ (Sect. \ref{raddenprof}), $m_2=1880\pm820 M_{\odot}$ (Sect.~8) and $m_3=11400\pm4550 M_{\odot}$ (Sect. \ref{dynamical}).

To characterise the mass loss due to the stellar evolution (equation (2) in Dalessandro et al. (2015)), we use the metallicity $Z=0.02$ (the coefficients given in Table\,1 of Lamers et al. (2005), which is sufficiently close to the value of $0.025$ estimated here.
The tidal mass loss is characterised by the dissolution timescale $t_0=3.3^{+1.4}_{-1.0}$\,Myr.
This value was derived for the solar neighbourhood and is a good approximation for the cluster NGC\,4337 that currently located at the Galactocentric radius of $\sim7.8\pm0.1$\,kpc.
Finally, for the coefficient $\gamma$ characterising the initial density distribution we adopt $\gamma=0.62$ which is a typical value for open clusters (Baumgardt \& Makino 2003)

We consider the uncertainty intervals for the current cluster age $t$, the mass estimates $m_{1,2,3}$, and the dissolution timescale $t_0$, which has the largest influence on the initial mass estimate
(Dalessandro et al. 2015).
This leads to the initial masses of about
$m_{\rm{ini},1}\approx$20--24$\times10^3$\,M$_{\odot}$,
$m_{\rm{ini},2}\approx$19--24$\times10^3$\,M$_{\odot}$,
$m_{\rm{ini},3}\approx$35--53$\times10^3$\,M$_{\odot}$.
We note that these values represent a rough estimate and are based on an approximate method and parameters derived for average open clusters observed in the solar neighbourhood.

In the most reliable case of the actual mass estimate that includes remnants and low mass stars, the cluster would have lost  between 60$\%$ and 80$\%$ of its initial mass.

\section*{Acknowledgments}
The work of A.F.Seleznev and the science leave of G.Carraro in Ekaterinburg was supported
by Act 211 Government of the Russian Federation, contract No. 02.A03.21.0006 and by the ESO DGDF
program. G. Baume acknowledges support from CONICET (PIP 112-201101-00301) and
financial support from the ESO visitor program that allowed a visit to ESO
premises in Chile, where part of this work was done.
This publication also made use of data from the Two Micron All Sky Survey,
which is a joint project of the University of Massachusetts and the Infrared
Processing and Analysis Center/California Institute of Technology, funded by
the National Aeronautics and Space Administration and the National Science
FoundationWe thank the Centre de Donn\'{e}es Astronomiques de Strasbourg (CDS), the U. S. Naval Observatory and NASA for the use of their electronic facilities, especially SIMBAD, ViZier and ADS, and the WEBDA database.



\end{document}